\documentclass[traditabstract]{aa} % for a referee version
\usepackage{graphicx}
\usepackage{rotating}
\usepackage{natbib}
\usepackage[usenames]{color}
\usepackage{multirow}
\usepackage{url}
\usepackage{soul}
\bibpunct{(}{)}{;}{a}{}{,} % to follow the A&A style
\newcommand{\ratio} {N({\rm H}_2) / I_{\rm CO(1-0)}}
\newcommand{\ratioo} {N({\rm H}_2) / I_{\rm CO}}
\newcommand{\ratiot} {N({\rm H}_2) / I_{\rm CO(2-1)}}
\def \ratioc{N({\rm H}_2) / I_{\rm \ion{C}{I}}}
\newcommand{\kms}   {{\rm \  km \  s^{-1}}}
\newcommand{\NHI}  {N(\ion{H}{I})}
\newcommand{\NHmol} {N(\mathrm{H_{2}})}
\newcommand{\X} {\NHmol/\mathrm{I}_{\mathrm{CO}}}

\newcommand{\Xunit} {\,{\rm cm^{-2}/(K\kms)}}
\newcommand{\mum} {$\mu$m }

\usepackage{txfonts}
\begin{document}
   \title{The Molecular Interstellar Medium of the Local Group Dwarf NGC~6822}
   \titlerunning{The Molecular ISM of NGC~6822}
   \subtitle{The Molecular ISM of NGC~6822}

   \author{P. Gratier \inst{1} \and J. Braine \inst{1} \and N.J. Rodriguez-Fernandez   \inst{2} \and F.P. Israel \inst{3}   \and K.F. Schuster  \inst{2} \and N. Brouillet \inst{1}  \and E. Gardan \inst{1}         }

   \institute{Laboratoire d'Astrophysique de Bordeaux, Universit\'e de Bordeaux, OASU, CNRS/INSU, 33271 Floirac France\\
              \email{gratier@obs.u-bordeaux1.fr}
         \and
            IRAM, 300 Rue de la piscine, F-38406 St Martin d'H\`eres, France   
         \and
            Sterrewacht Leiden, PO Box 9513, 2300 RA Leiden, The Netherlands
        }

   \date{}

  \abstract{Do molecular clouds collapse to form stars at the same rate in all environments?
In large spiral galaxies, the rate of transformation of H$_2$ into stars (hereafter SFE) varies little.
However, the SFE in distant objects ($z \sim 1$) is much higher than in the large spiral disks that dominate the local universe.
   Some small local group galaxies share at least some of the characteristics of intermediate-redshift objects, such as size or color.
Recent work has suggested that the Star Formation Efficiency (SFE, defined as the SFRate per unit H$_2$) in local Dwarf galaxies may be as high as in the distant objects.
A fundamental difficulty in these studies is the independent measure of the H$_2$ mass in metal-deficient environments.
  At 490 kpc, NGC~6822 is an excellent choice for this study;
it has been mapped in the CO(2--1) line using the multibeam receiver HERA on the 30 meter IRAM telescope, yielding the  largest sample of giant molecular clouds (GMCs) in this galaxy.  Despite the much lower metallicity, we find no clear difference in the properties of the GMCs in NGC~6822 and those in the Milky Way except lower CO luminosities for a given mass.  Several independent methods indicate that the total H$_2$ mass in NGC~6822 is about $5 \times 10^6$ M$_{\sun}$ in the area we mapped and less than 10$^7$ M$_{\sun}$ in the whole galaxy.  This corresponds to a $\ratioo \approx 4 \times 10^{21}\Xunit$ over large scales, such as would be observed in distant objects, and half that in individual GMCs.  No evidence was found for H$_2$ without CO emission.  Our simulations of the radiative transfer in clouds are entirely compatible with these $\ratioo$ values.  The SFE implied is a factor 5 -- 10 higher than what is observed in large local universe spirals.  The CO observations presented here also provide a high-resolution datacube (1500~a.u. for the assumed 100~pc distance, $0.41\kms$ velocity resolution) of a local molecular cloud along the line of sight.}

\maketitle
%
%________________________________________________________________

\section{Introduction}
In the study of star formation in a cosmological or extragalactic context, rather than the details of the collapse of a cloud core to a star, we are interested in understanding why stars form where they do, whether the efficiency varies, and what factors influence the initial mass function (IMF) of the stars.  Over the last decade, it has become very clear that the star formation rate per co-moving volume was much higher in the past, some 10 or 20 times the current rate at a redshift of $z \sim 0.5 - 1$ \citep[e.g.][]{Madau:1996,Heavens:2004,Wilkins:2008}.  In turn, this shows that the transformation rate of gas into stars was considerably (factor few at least) higher when the universe was roughly half its current age.  Galaxies at that time were smaller and of lower metallicity, such that naively at least one would expect that the molecular-to-atomic gas mass ratio would be lower than today \citep{Young:1989,Casoli:1998}, making the higher efficiency even more surprising.
%It is generally accepted that the Star Formation Rate (SFR)
%increases very sharply in galaxies up to a redshift of about $z = 1$.
Since stars form from H$_2$, and not directly
from \ion{H}{I} (with the possible exception of the so-called Pop. III, or first
generation stars), this suggests that either large amounts of molecular hydrogen
were available or that for some reason the efficiency of star formation (SFE,
defined as the SFR per unit H$_2$) was particularly high back then.  In
fact, the SFRs proposed are so much higher than the SFR today that both possibilities may be
required.  Because at least 10\% of the baryons in galaxies today are thought
to be in  neutral gas (and more than 10\% in many cases), an SFR a factor
15 -- 20 higher must result at least partially from a higher SFE.
If the SFE is higher, then something about the process of star formation is different
and there could be other important differences like a change in IMF.
%While there may not be a 100\% consensus, it is widely believed that the
%fraction of molecular gas (with respect to atomic) decreases going to later types and smaller objects.
Moderate to high redshift galaxies are typically
smaller and more gas-rich than today's spirals and most likely have a slightly
subsolar metallicity.  They thus resemble today's small spirals such as M33,
or the even smaller NGC 6822,
and could be expected to have a low H$_2$/\ion{H}{I} mass ratio.
If so, this would make the SFE in these objects even more extreme.

The first step is to learn more about the molecular gas content of galaxies
with these properties. Significant quantities of molecular gas were detected far out
in the outer disks of NGC~4414 and NGC~6946 \citep{Braine:2004, Braine:2007}.
The outskirts of spirals share the subsolar metallicities and low
mass surface densities of small and/or medium/high redshift spirals but not the
level of star formation.  We are fortunate to have a number of small galaxies in the Local Group
close enough that individual giant molecular clouds (GMCs) can be resolved and without distance
ambiguities.  A first step has been taken, showing that
molecular gas forms very far out in M~33 despite the low metallicity and very low
ambient pressure \citep{Gardan:2007}.
NGC 6822 is among the nearest galaxies and is a small late-type dwarf spiral at a distance of about 490 kpc
\citep{Mateo:1998} and has a mass and luminosity
of roughly 1\% of that of our galaxy, thus representing a step down in mass,
luminosity and metallicity (roughly $12+log(O/H) \sim 8.1$ \citep{Lee:2006})  with respect to M33,
itself a step down in the same quantities from the Milky Way or M31.
At 490 kpc, $1\arcsec$ corresponds to 2.4 pc, such that Giant Molecular Clouds (GMCs)
can be resolved with large single-dish radiotelescopes.

Observations of the universe at redshifts $z \ga 1$ show that todays large
spirals were not present or rare at these earlier epochs.  Rather, the galaxies
were smaller and had higher star formation rates.  They probably had somewhat lower metallicities and were bluer.  M33 was found to have a high SFE \citep{Gardan:2007},
compared to local universe spirals and IC10 appears
to show a high SFE as well \citep{Leroy:2006}.  NGC 6822 is closer and its molecular gas content
has not been mapped systematically until now. 
Giant Molecular Clouds have been resolved and their physical properties studied in Local Group galaxies both with single dish telescopes (M33 by \citealt{Gardan:2007}; SMC by \citealt{Rubio:1993}; LMC by \citealt{Fukui:2008}; SMC and LMC by \citealt{Israel:2003a}) and with interferometers (M33 by \citealt{Engargiola:2003}; IC~10 by \citealt{Leroy:2006}; M31 by \citealt{Rosolowsky:2007a}).
NGC 6822 has been observed at many wavelengths to study the interstellar medium (ISM),
the dynamics, and trace the star formation.
Spitzer FIR observations were carried out recently by \citet{Cannon:2006}. The
atomic gas has been mapped \citep{de-Blok:2000, de-Blok:2003, de-Blok:2006, de-Blok:2006a,Weldrake:2003}
and the molecular gas observed at specific positions with the 15m SEST and JCMT telescope \citep{Israel:1996, Israel:1997, Israel:2003}.With the OVRO interferometer \citep{Wilson:1994} observed 3 GMCs in NGC~6822 in the Hub V region.
Since one of the main questions is whether significant quantities of molecular gas
could be present without detectable CO emission, high sensitivity high resolution mapping of large regions allowing the detections of individual possibly optically thick clouds is required. 
The molecular gas content derived via CO can then be compared with other means of
tracing the molecular and atomic gas.   

In this article we present the observations and data reduction, mostly of
CO (but also \element[][13]{C}O and HCN(1--0) in Hubble V), followed by the production of a
catalog of molecular clouds and their properties which we compare with Galactic GMCs.
Two methods were used to compile the catalog of cloud sizes, CO intensities, and virial masses: visual inspection of the data cube and the {\tt CPROPS} algorithm \citep{Rosolowsky:2006}.
A map of the total CO emission is then compared (Sect. \ref{sec.mass}) with other means of estimating
the H$_2$ column density, leading to a discussion of the SFE in NGC~6822.
We then present several models using the CLOUDY \citep{Ferland:1998} code to compare several spectral synthesis models with the observations presented in the preceding sections.
Finally, two regions of NGC~6822 are discussed in more detail -- Hubble V and X.
A local molecular cloud is present along the line of sight to NGC~6822 and has thus
been observed serendipitously at high spectral and spatial resolution.  These data are
presented in the last section.

\begin{figure*}[t]
\begin{flushleft}
\includegraphics[angle=270,width=75cm]{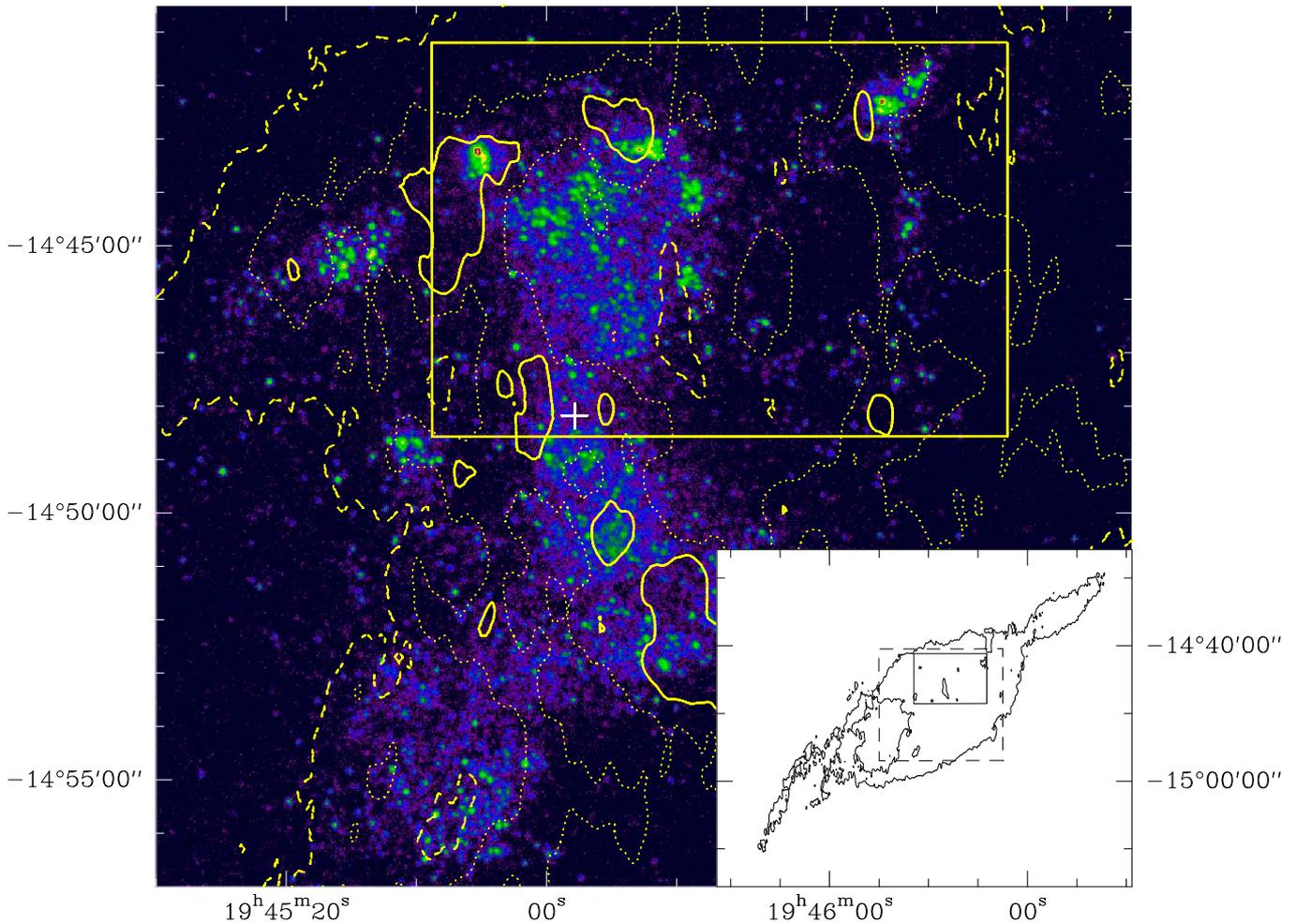}
\caption{\emph{Main image} GALEX FUV filter image of NGC~6822.  
The yellow box indicates the area observed in CO.
The dashed, dotted, and solid yellow contours indicate the 
\ion{H}{I} column density at $12\arcsec \times\ 42\arcsec$ resolution from \citet{de-Blok:2006a}
at column densities of 0.5, 1, and 1.5 $\times\ 10^{21}$ cm$^{-2}$ respectively. The white cross corresponds to (0,0) offsets of the HERA CO(2-1) map. \emph{Bottom right} HI contour at a column density of $0.5\times10^{21}$ cm$^{-2}$. The dashed rectangle corresponds to the GALEX FUV map above, the solid line contour to the area observed in CO.}
\label{fig.galex}
\end{flushleft}
\end{figure*}

\begin{table}[tbp]
\caption{Parameters for NGC 6822.}
\label{tab.param}
%\begin{flushleft}
\begin{tabular*}{88mm}{@{\extracolsep{\fill}}l  r}
\hline
\hline
$\alpha_{o}$(J2000)$^{\mathrm{a}}$ & $19^\mathrm{h}44^\mathrm{m}57\ \fs83$\\
$\delta_{o}$(J2000)$^{\mathrm{a}}$ & $-14\degr48\arcmin11\arcsec $\\

Distance$^{\mathrm{b}}$ & $490\pm40$ kpc\\
$V_{\mathrm{LSR}}$ & $-47$ $\kms$\\
Metallicity$^{\mathrm{c}}$ & $\simeq 0.3~\mathrm{Z}_{\sun} $\\
Total optical luminosity$^{\mathrm{b}}$ & $\sim 9.4 \times 10^7~\mathrm{L}_{\sun}$\\
Total \ion{H}{I} mass$^{\mathrm{d}}$ & $1.34\times10^8~\mathrm{M}_{\sun}$\\
Dynamical mass$^{\mathrm{e}}$ & $3.2 \times10^9~\mathrm{M}_{\sun}$\\
\hline
\end{tabular*}
%\begin{flushleft}
\begin{list}{}{}
\item[$^{\mathrm{a}}$] (0,0) offset of our HERA CO(2-1) map
\item[$^{\mathrm{b}}$] \citet{Mateo:1998}
\item[$^{\mathrm{c}}$] \citet{Skillman:1989,Lee:2006}
\item[$^{\mathrm{d}}$] \citet{de-Blok:2006a}
\item[$^{\mathrm{e}}$] \citet{de-Blok:2006a,Weldrake:2003}
\end{list}
\end{table}
 
 \section{Observations}

NGC~6822 was observed during three separate runs at the IRAM 30 meter telescope in
November 2006, February/March 2007 and August 2008 in mostly good weather. All mapping was done using the HERA array of 9 dual polarization
receivers \citep{Schuster:2004} in the CO(2--1) line, whose rest frequency
is 230.53799 GHz which gives a nominal resolution of $11.4\arcsec$ for this line. The On-The-Fly mode was used to cover
a roughly $11 \times 7$ arcminute  region, scanning along the RA
and then Dec directions and observing a reference position
offset from our central position by $(100\arcsec, 500\arcsec)$
before and after each scan.  The reference position was chosen to be
outside of the $5 \times 10^{20}$ cm$^{-2}$ contour of the \ion{H}{I}
column density map (see Fig. \ref{fig.galex}).

The VESPA backend was used with a channel spacing of 312 kHz
or $0.406\kms$ covering velocities from $+20\mbox{ to }{-150\kms}$, well beyond
the rotation curve of NGC~6822.  Local (Galactic) emission was detected
around ${+5\kms}$ (see Sect. \ref{sec.cirrus}).  All data are presented in the main beam temperature
scale and we have assumed forward and main beam efficiencies of $\eta_{for} = 0.90$
and $\eta_{mb} = 0.52$ for the HERA observations, the sensitivity is then ${\rm 9.6 Jy/K}$ \citep{Schuster:2004}.

Also during the November 2006 run, during poorer weather than for the
more demanding HERA observations, the major \ion{H}{II} region Hubble V was observed in
the \element[][13]{C}O(1--0), \element[][13]{C}O(2--1), HCN(1--0), and \element[][12]{C}O(1--0) lines.
Wobbler switching was used with a throw of $60\arcsec$
and the 100 kHz and VESPA backends were used, yielding spectral resolutions of
respectively 0.27, 0.43, 1.06, and 0.26 $\kms$ for the lines above.
The forward and main beam efficiencies at these frequencies are
assumed to be respectively $\eta_{for} = 0.95, \  0.91, \  0.95, \  0.95$ and
$\eta_{mb} = 0.75, \  0.55, \  0.78, \  0.74$.
The data reduction of the HERA observations is described in the next section.
For the Hubble V data, bad channels were eliminated and spectra were averaged,
yielding the spectra discused in Sect. \ref{sec.HubV}.

 \section{Reduction of HERA data}
 
 The On-The-Fly mapping technique with a multi-beam array generates a huge
 amount of spectra, more than a million in the case of NGC~6822.  Inspection of individual
 spectra is thus not possible and the reduction was automated.
 All data reduction was done within the Gildas\footnote{\url{http://www.iram.fr/IRAMFR/GILDAS}} CLASS and GREG software packages.
 After filtering out the spectra taken in very poor conditions (T$_{sys} > 1500$~K),
 we treated the main problem which was the slight platforming where the
 sub-bands of the auto-correlator backend were stitched together.
The platforming effect is the result of different non-linearities in the sampling stages of the subbands. In case of changing total power levels as compared to the reference position this introduces offsets in the subbands.  The steps were very small in our case
 but sufficient to affect weak lines.
 For each spectrum, the average value for each subband, outside of the line windows as far as possible, was
 subtracted from each channel of the sub-band, eliminating the platforming.
 This process takes out a zero-order baseline.
 
 There is a known baseline ripple with HERA \citep{Schuster:2004} corresponding
 to a reflection off the secondary mirror at 6.9 MHz or 9 km s$^{-1}$.
 Since the pixels (receivers) are affected at quite different levels and because the 
 width of the ripple is close to that of molecular clouds, we tested each spectrum
 and when the ripple was strong enough to be identified in the individual (i.e. roughly 
 1sec integration time) spectra, the corresponding Fourier frequency was replaced
 by an interpolation based on adjacent frequencies.  This fairly standard filtering is often
 applied ``blindly'' to all spectra but we only applied it when necessary.
 
In order to create the datacube, we created a data table with the TABLE 
command within CLASS90 and then used the {\sc XY\_MAP} task in GREG with parameters
such that the final resolution became $15\arcsec$, or about 36 pc at the distance of NGC~6822.
Data cubes with different resolutions where always generated directly from the original data by convolution with the corresponding kernel size. 
\section{The individual molecular clouds in NGC~6822} 

\label{sec.clouds}

\subsection{Identification by visual inspection}
\label{sec.cloudID}
\begin{figure*}[tp]
\begin{flushleft}
\includegraphics[angle=270,width=180mm]{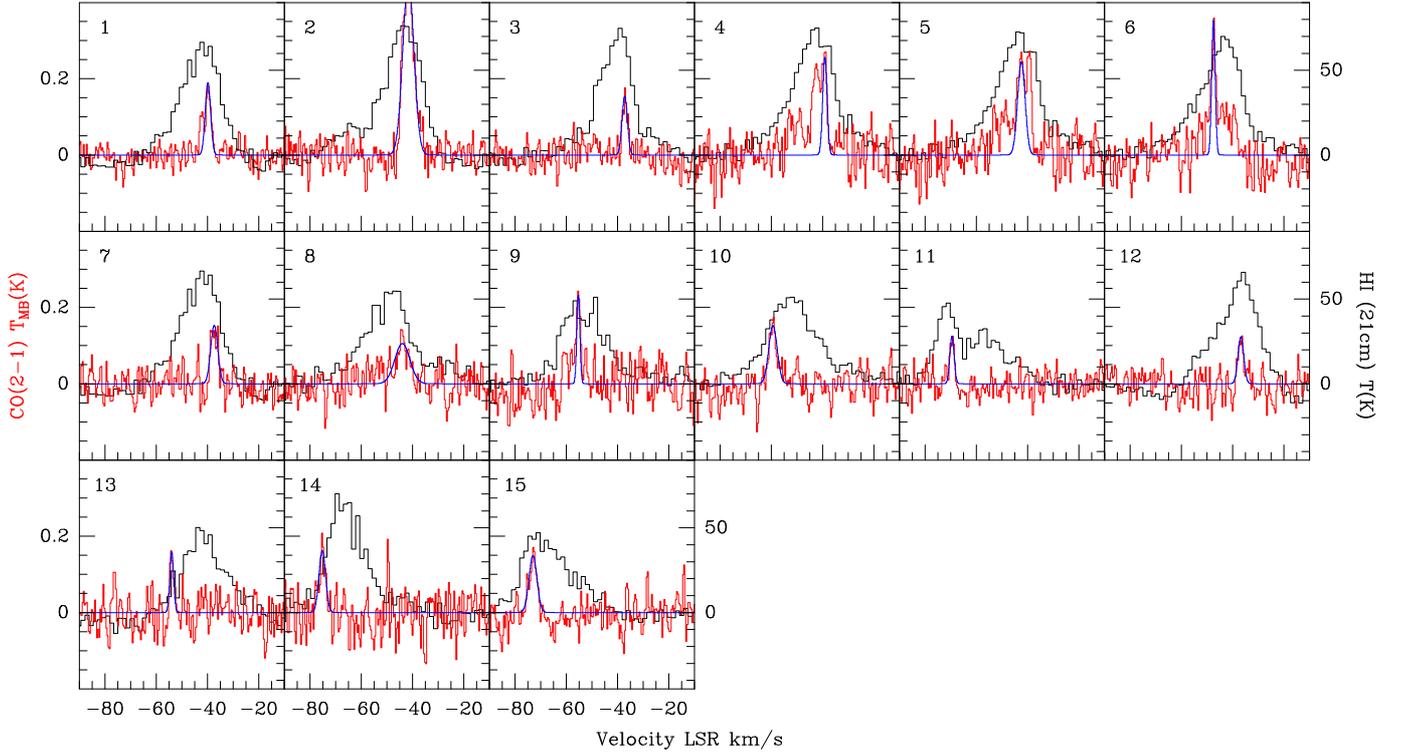}
\caption{Spectra of individual clouds, (\emph{black}) \ion{H}{I} 21cm line, (\emph{red}) CO(2-1) line, (\emph{blue}) Gausian line fit to the narrowest component, the physical parameters of these clouds can be found in Table \ref{tab.clouds}.}
\label{fig.spectra}
\end{flushleft}
\end{figure*}

The final beam width of the CO observations is about 30 pc and because of the beam dilution we do not expect to see clouds with sizes under 10 pc. We define clouds as structures similar to Galactic Giant Molecular Clouds (GMCs) that appear as gravitationally bound and non transient structures a few ten of parsecs in size.
Figure \ref{fig.spectra} shows the CO(2--1) and \ion{H}{I} spectra of the clouds found in the original $15\arcsec$ datacube, showing
both the CO and \ion{H}{I} intensity scales. The spectra are averages over the 50\% brightness contour.  It is clear that all strong CO lines are close to the \ion{H}{I} peak in velocity.
However, cloud 13 is at either the edge of the \ion{H}{I} line or possibly part of a second \ion{H}{I} feature with a brightness
of 25~K.  Cloud 14 is also near the edge of the \ion{H}{I} line although mostly within the 30~K \ion{H}{I} brightness temperature
level used to define the CO line window in the next section.

Gaussian fits were made to the individual clouds in order to determine linewidths and central velocities.
In several cases, more than one gaussian was required. Since one of the goals is to measure line width and cloud sizes for NGC~6822 clouds and compare to Milky Way clouds, only the stronger (in antenna temperature) and narrower gaussian
was used to define the line widths and sizes of the individual clouds. Line areas are computed by summing channels in a velocity range determined manually for each cloud. The line intensities, systemic velocities and widths are averages over the 50\% brightness level of each cloud.

\begin{figure}[tp]
\begin{flushleft}
\includegraphics[angle=270,width=88mm]{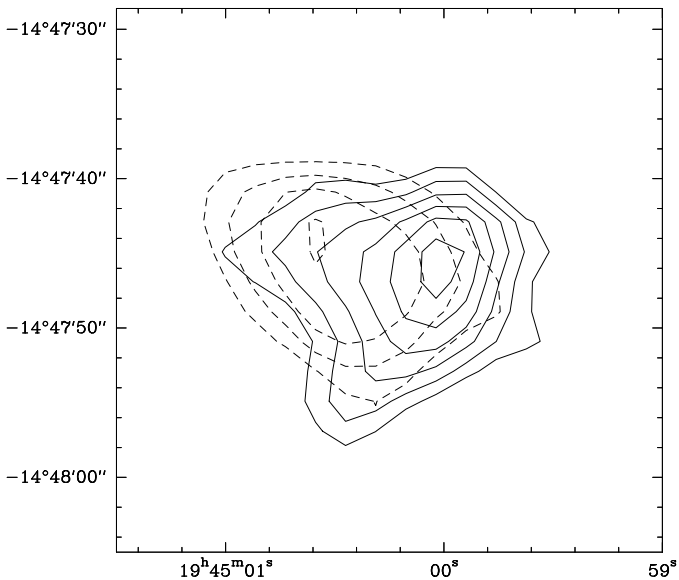}
\caption{Contour integrated intensity maps of clouds 4 (solid lines)  and 5 (dashed lines) contours every $0.1 \rm{K}\kms$.  The emission peaks are separated by about 9 arcsec
showing that there are clearly two separate clouds.}
\label{fig.clouds_2and3}
\end{flushleft}
\end{figure}

Spectra for clouds 4 and 5 show that these clouds are only partially spatially resolved. Contour maps of the integrated intensity for the two components 4 and 5 (Fig. \ref{fig.clouds_2and3}) indicate that the emission can be separated into two clouds separated by about 9 arcseconds.

Table 2 provides positions, velocities, linewidths, and estimated sizes for each cloud.
In some cases, clouds were separated by summing over different velocity ranges,
enabling cloud separations smaller than the beamsize.
On the individual maps, corresponding to each cloud, the 50\% brightness level was defined for each cloud,
allowing the effective radius (see Sect. \ref{sec.CPROPS})  of the cloud to be estimated as $R_{e} = (1.92/2\sqrt{2\ln2})  \sqrt{polygon\ area\arcsec - 15\arcsec^2}$  and converted to pc.  
The subtraction of the beam area ($15^2$) enables a simple deconvolution with the the beamsize assuming in average a gaussian intensity distribution of the cloud emission, the numerical factor converts the FWHM to an effective radius. Below about 10 pc, the beam dilution makes the detection of individual clouds difficult. 

The CO(2--1) integrated intensities (col. 7) are obtained by summing channels within $V=V_{o}\pm \Delta V$.
Virial masses (col. 8) are calculated from $M_{vir} = 1040R_{e}\left({\Delta V}/{2\sqrt{2\ln2}}\right)^2$ following \citet{Solomon:1987} and using their
form factor of 2.7.  
For comparison, cloud masses (including He) can be estimated as
$M = I_{\rm CO(2-1)} \times 2500 \times ({\rm size/20pc})^2$M$_{\sun}$
using a ``standard" Milky Way factor of
$\ratio = 2 \times 10^{20}\Xunit$  \citep{Dickman:1986} and a CO(2--1)/(1--0) line ratio of 0.7 \citep{Sawada:2001}.
Inspection of the values shows that the Galactic $\ratioo$ factor yields masses far below
the virial masses. This is well known for clouds in low metallicity galaxies \citep[e.g.][]{Rubio:1993,Israel:1997a}.  The virial masses themselves may be underestimates of the true H$_2$ masses
if the H$_2$ extends beyond where the CO is detected.  The following column provides an independent 
estimate of the H$_2$ mass (see Sect. \ref{sec.mass8mu} for details) and resulting $\ratioo$ factor.

\begin{table*}
\begin{flushleft}
\caption{Properties for the clouds identified by eye (\emph{Top}), and by the {\tt CPROPS} package (\emph{Bottom}).}
The clouds whose numbers are in bold face correspond to our final sample of clouds\\

\label{tab.clouds}
%A la main
{\renewcommand{\arraystretch}{1.2}
\begin{tabular*}{180mm}{@{\extracolsep{\fill}}lrrrrrrrrr}

\hline 
\hline 
Cloud & $\left.\alpha_{\mathrm{off}}\right.^{\mathrm{a}}$ & $\left.\delta_{\mathrm{off}}\right.^{\mathrm{a}}$ & $\mathrm{V}_{\mathrm{LSR}}$ & \mbox{\textsc{FWHM}} & ${\rm R_e}$ & $\mathrm{I}_{\mathrm{CO}} $ & $\left.\mathrm{M}_{\mathrm{vir}}\right.^{\mathrm{b}}$ & $\left.\mathrm{M}_{\mathrm{H_{2}}}\right.^{\mathrm{c}}$ &$\left.\X\right.^{\mathrm{d}}$\\ 
& arsec & arcsec & $\kms$ & $\kms$ & pc & $\mathrm{K}\kms$ & $ \mathrm{M}_{\sun} $ & $ \mathrm{M}_{\sun}$ & $ \frac{\mathrm{cm}^{-2}}{\mathrm{K}\kms}$\\[0.5em] 
\hline
%\rule{0pt}{2ex}
{\bf GMC NGC6822\,1}&$-48$& 330&$   -43.2$&$    1.5\pm 0.9$&$22_{-14}^{+9}$&$0.92_{-0.10}^{+0.10}$&$9.3_{-8.7}^{+24}\times 10^3$&$3.6_{-0.8}^{+1.4}\times 10^4$&$1.4_{-0.2}^{+0.2}\times 10^{21}$\\
 GMC NGC6822\,$2^{\mathrm{e}}$&  $ -68.5$&   299&$   -42.4$&$    5.1\pm 0.4$&$29_{-6}^{+5}$&$2.47_{-0.13}^{+0.14}$&$1.9_{-0.5}^{+0.5}\times 10^5$&$3.1_{-0.6}^{+0.3}\times 10^4$&$3.0_{-0.3}^{+0.3}\times 10^{20}$\\
 GMC NGC6822\,3&    80&    58&$   -38.0$&$    2.4\pm 0.5$&    $38_{-21}^{+15}$&    $0.36_{-0.06}^{+0.07}$&$       4.1_{-3}^{+4.2}\times 10^4$&$       9.5_{-3.7}^{+3.5}\times 10^4$&$ 4.3_{-1.1}^{+1.6}\times 10^{21}$\\
 GMC NGC6822\,4&    39&    25&$   -39.9$&$    2.0\pm 0.6$&$18_{-10}^{+6}$&$1.34_{-0.09}^{+0.10}$&$1.2_{-0.3}^{+0.1}\times 10^5$&$ 1.2_{-0.3}^{+0.1}\times 10^5$&$ 3.1_{-0.3}^{+0.4}\times 10^{21}$\\
{\bf GMC NGC6822\,5}&    35&    24&$   -43.2$&$    3.4\pm 1.1$&$19_{-9}^{+8}$&$1.52_{-0.13}^{+0.11}$&$ 4.1_{-0.3}^{+0.6}\times 10^4$&$ 1.2_{-0.3}^{+0.2}\times 10^5$&$ 2.5_{-0.2}^{+0.5}\times 10^{21}$\\
 GMC NGC6822\,6&    30&    16&$   -48.3$&$    1.6\pm 0.4$&    $12_{\textrm{ \ldots}}^{+15}$&    $0.82_{-0.07}^{+0.05}$&$       5.8_{\textrm{ \ldots}}^{+14}\times 10^3$&$ 4.7_{-1.4}^{+1.4}\times 10^4$&$ 2.4_{-0.2}^{+0.3}\times 10^{21}$\\
 GMC NGC6822\,7& $ -10$&  364&$   -38.1$&$    3.2\pm 0.4$&$27_{-15}^{+11}$&$ 0.48_{-0.07}^{+0.06}$&$5.2_{-3.4}^{+4.1}\times 10^4$&$4.9_{-1.0}^{+0.2}\times 10^3$&$2.8_{-1.2}^{+1.6}\times 10^{20}$\\
{\bf GMC NGC6822\,8}& 31&   61&$   -44.8$&$    2.4\pm 1.4$&    $14_{\textrm{ \ldots}}^{+28}$&$0.58_{-0.14}^{+0.13}$&$ 1.5_{\textrm{ \ldots}}^{+9.9}\times 10^4$&$4.1_{-2.5}^{+5.1}\times 10^4$&$2.8_{-0.5}^{+1.0}\times 10^{21}$\\
 GMC NGC6822\,9&     7&    54&$   -56.0$&$    1.4\pm 0.4$&$\textrm{ \ldots}_{\textrm{ \ldots}}^{+13}    $&$0.57_{-0.10}^{+0.04}$&$\textrm{ \ldots}_{\textrm{ \ldots}}^{+7.9}\times 10^3$&$7.6_{-1.6}^{+4.2}\times 10^3$&$8.2_{-0.7}^{+1.7}\times 10^{20}$\\
GMC NGC6822\,10&     0&    26&$   -60.1$&$    3.9\pm 1.0$&$33_{-17}^{+17}$&    $0.59_{-0.10}^{+0.07}$&$       9.4_{-6.9}^{+13}\times 10^4$&$       7.6_{-2.9}^{+3.4}\times 10^4$&$2.6_{-0.7}^{+1.0}\times 10^{21}$\\
GMC NGC6822\,11& $-358$&  208&$   -70.3$&$    2.0\pm 0.9$&$28_{\textrm{ \ldots}}^{43}$&$0.23_{-0.07}^{+0.07}$&$ 2.1_{\textrm{ \ldots}}^{+9.1}\times 10^4$&$\textrm{ \ldots}_{\textrm{ \ldots}}^{\textrm{ \ldots}}\,^{\mathrm{h}}$&$\textrm{ \ldots}_{\textrm{ \ldots}}^{\textrm{ \ldots}}\,^{\mathrm{h}}$\\
\hline  
{\bf GMC NGC6822\,12}&$ 55$&$ 287$&$-37.6$&$2.6\pm 0.5$&$23_{\textrm{ \ldots}}^{+28}$&$0.23_{-0.02}^{+0.06}$&$2.9_{\textrm{ \ldots}}^{+6.3}\times 10^4$&$\textrm{ \ldots}_{\textrm{ \ldots}}^{+3.0}\times 10^3$&$\textrm{ \ldots}_{\textrm{ \ldots}}^{+1.7}\times 10^{20}$\\
{\bf GMC NGC6822\,13}&$    49$&$   364$&$   -54.7$&$    1.7\pm 0.4$&$\textrm{ \ldots}_{\textrm{ \ldots}}^{+34}$&$0.26_{-0.07}^{+0.05}$&$\textrm{ \ldots}_{\textrm{ \ldots}}^{+2.7}\times 10^4$&$\textrm{ \ldots}_{\textrm{ \ldots}}^{\textrm{ \ldots}}\,^{\mathrm{h}}$&$\textrm{ \ldots}_{\textrm{ \ldots}}^{\textrm{ \ldots}}\,^{\mathrm{h}}$\\
% {14}&$    36$&$   317$&$   -43.1$&$    1.5\pm 0.5$&$14_{\textrm{ \ldots}}^{+28}$&$0.23_{-0.08}^{+0.03}$&$5.9_{\textrm{ \ldots}}^{+26}\times 10^3$&$\textrm{ \ldots}_{\textrm{ \ldots}}^{\textrm{ \ldots}}$&$\textrm{ \ldots}_{\textrm{ \ldots}}^{\textrm{ \ldots}}$\\
% {15}&$    36$&$   317$&$   -40.3$&$    2.3\pm 0.4$&$14_{\textrm{ \ldots}}^{+28}$&$0.54_{-0.10}^{+0.07}$&$1.4_{\textrm{ \ldots}}^{+4.4}\times 10^4$&$\textrm{ \ldots}_{\textrm{ \ldots}}^{\textrm{ \ldots}}$&$\textrm{ \ldots}_{\textrm{ \ldots}}^{\textrm{ \ldots}}$\\
{\bf GMC NGC6822\,14}&$  -331$&$   372$&$   -76.0$&$    3.0\pm 1.0$&$\textrm{ \ldots}_{\textrm{ \ldots}}^{+15}$&$0.32_{-0.08}^{+0.08}$&$\textrm{ \ldots}_{\textrm{ \ldots}}^{+4.5}\times 10^4$&$\textrm{ \ldots}_{\textrm{ \ldots}}^{\textrm{ \ldots}}\,^{\mathrm{h}}$&$\textrm{ \ldots}_{\textrm{ \ldots}}^{\textrm{ \ldots}}\,^{\mathrm{h}}$\\
{\bf GMC NGC6822\,15}&$  -386$&$    57$&$   -73.8$&$    3.6\pm 0.8 $&$24_{-16}^{+21}$&$0.53_{-0.12}^{+0.06}$&$5.8_{-4.7}^{+11}\times 10^4$&$5.5_{-1.4}^{+1.3}\times 10^4$&$2.9_{-01.2}^{+1.7}\times 10^{21}$\\
%$ {14\mbox{\&}15}\ ^{\mathrm{f}}$&$  36$&$    317.1$&$   -40.9$&$    3.7\pm0.4$&$14_{\textrm{ \ldots}}^{+28}$&$0.77_{-0.05}^{+0.09}$&$3.6_{\textrm{ \ldots}}^{+9.6}\times 10^4$&$\textrm{ \ldots}_{\textrm{ \ldots}}^{\textrm{ \ldots}}$&$\textrm{ \ldots}_{\textrm{ \ldots}}^{\textrm{ \ldots}}$\\
\hline 
\end{tabular*}}

\vspace{5mm}

%Avec CPROPS
{\renewcommand{\arraystretch}{1.2}
\begin{tabular*}{180mm}{@{\extracolsep{\fill}}lrrrrrrrrrr} 
\hline 
\hline 
Cloud & $\left.\alpha_{\mathrm{off}}\right.^{\mathrm{a}}$ & $\left.\delta_{\mathrm{off}}\right.^{\mathrm{a}}$ & $\mathrm{V}_{\mathrm{LSR}}$ & \mbox{\textsc{FWHM}} & ${\rm R_e}$ & ${\rm R_{ex}}$ & $\mathrm{I}_{\mathrm{CO}} $ & $\left.{\mathrm{M}_{\mathrm{vir}}}\right.^{\mathrm{b}}$ & $\left.\mathrm{M}_{\mathrm{H_{2}}}\right.^{\mathrm{c}}$ &$\left.\X\right.^{\mathrm{d}}$\\ 
& arsec & arcsec & $\kms$ & $\kms$ & pc & pc & $\mathrm{K}\kms$ & $ \mathrm{M}_{\sun} $ & $ \mathrm{M}_{\sun}$ & $ \frac{\mathrm{cm}^{-2}}{\mathrm{K}\kms}$\\[0.5em] 
\hline 
%\rule{0pt}{2ex}
GMC NGC6822\,1&$    -50$&$    331$&$   -40.2$&$     5.5\pm     3.5^{\mathrm{g}}$&$     15\pm     14$&$     38$&$     0.9\pm     0.3$&$       8.5\pm       9.2\times 10^4 \,^{\mathrm{g}}$&$       5.1\times 10^4$&$       9.9\times 10^{20}$\\
{\bf GMC NGC6822\,2}&$    -71$&$    301$&$   -41.3$&$     6.2\pm     1.4$&$     32\pm      4$&$     86$&$     1.2\pm     0.1$&$       2.3\pm       0.5\times 10^4$&$       9.8\times 10^4$&$       4.2\times 10^{20}$\\
{\bf GMC NGC6822\,3}&$     83$&$     59$&$   -37.0$&$     2.2\pm     1.9$&$     32\pm     13$&$     54$&$     0.2\pm     0.1$&$       2.8\pm       2.6\times 10^4$&$       1.1\times 10^5$&$       6.4\times 10^{21}$\\
{\bf GMC NGC6822\,4}&$     38$&$     25$&$   -39.2$&$     2.1\pm     3.2$&$     15\pm     21$&$     12$&$     1.0\pm     1.3$&$       1.3\pm       2.3\times 10^4$&$       1.2\times 10^5$&$       4.0\times 10^{21}$\\
GMC NGC6822\,5&$     35$&$     24$&$   -42.6$&$     3.9\pm     6.2$&$     10\pm     22$&$     \ldots^{\mathrm{\,i}}$&$     1.4\pm     2.2$&$       2.9\pm       7.6\times 10^4$&$       8.8\times 10^4$&$       2.8\times 10^{21}$\\
{\bf GMC NGC6822\,6}&$     30$&$     17$&$   -47.1$&$     3.2\pm     3.0$&$     15\pm     29$&$     33$&$     0.7\pm     0.2$&$       2.8\pm       6.3\times 10^4$&$       9.9\times 10^4$&$       3.0\times 10^{21}$\\
{\bf GMC NGC6822\,7}&$    -17$&$    359$&$   -38.1$&$     2.1\pm     1.8$&$     31\pm     17$&$     47$&$     0.3\pm     0.2$&$       2.5\pm       2.4\times 10^4$&$\ldots^{\mathrm{h}} $&$\ldots^{\mathrm{h}}$\\
GMC NGC6822\,8&$     33$&$     62$&$   -43.2$&$     4.6\pm     6.0$&$     19\pm     20$&$     37$&$     0.3\pm     0.2$&$       7.5\pm       14\times 10^4$&$       9.5\times 10^4$&$       4.7\times 10^{21}$\\
{\bf GMC NGC6822\,9}&$      7$&$     54$&$   -55.3$&$     2.1\pm     1.8$&$     10\pm     31$&$     31$&$     0.3\pm     0.1$&$       0.8\pm       2.3\times 10^4$&$       3.6\times 10^4$&$       2.6\times 10^{21}$\\
{\bf GMC NGC6822\,10}&$      0$&$     27$&$   -59.4$&$     4.6\pm     2.7$&$     25\pm      7$&$     55$&$     0.5\pm     0.1$&$       9.6\pm       6.1\times 10^4$&$       1.4\times 10^5$&$       3.1\times 10^{21}$\\
{\bf GMC NGC6822\,11}&$   -365$&$    210$&$   -69.4$&$     1.7\pm     1.8$&$     11\pm     27$&$     16$&$     0.2\pm     0.2$&$       0.6\pm       1.7\times 10^4$&$\ldots^{\mathrm{h}}$&$\ldots^{\mathrm{h}}$\\
\hline 
\end{tabular*}}
\end{flushleft}

\begin{list}{}{}
\item[$^{\mathrm{a}}$] Offsets with respect to the reference position $(\alpha_{o},\delta_{o})=(19^\mathrm{h}44^\mathrm{m}57\ \fs8,-14\degr48\arcmin11\arcsec )$
\item[$^{\mathrm{b}}$] Virial mass following \citet{Solomon:1987} using ${\rm R_e}$ and $\Delta V$ (see Sect. \ref{sec.clouds})
\item[$^{\mathrm{c}}$] $\mathrm{H}_{2}$ mass estimate from ${\rm 8\mu m}$ emission (see Sect. \ref{sec.mass8mu}).
\item[$^{\mathrm{d}}$] X conversion factor for individual clouds using $\NHmol$ estimated from ${\rm 8\mu m}$ emission (see Sect. \ref{sec.mass8mu}). \item[$^{\mathrm{e}}$] Hubble V \citep{Hubble:1925}
\item[$^{\mathrm{f}}$] see Sect. \ref{sec.clouds}
\item[$^{\mathrm{g}}$] The {\tt CPROPS} computed FWHM is clearly overestimated compared to the one computed by hand by a factor of about 3. ${\rm M_{vir}}$ being proportional to ${\rm FWHM^2}$, this explains the order of magnitude difference between the computed virial masses for this cloud.
\item[$^{\mathrm{h}}$] The computed ${\rm H_{2}}$ mass for these clouds was found to be negative.
\item[$^{\mathrm{\,i}}$] The deconvolved size was not defined.
\end{list}

\end{table*}

\subsection{Automated cloud identification}
\subsubsection{{\tt CPROPS}}
\label{sec.CPROPS}
We have also used the {\tt CPROPS}\footnote{\url{http://www.cfa.harvard.edu/~erosolow/CPROPS/}} program \citep{Rosolowsky:2006} to identify GMCs and measure their physical properties in an unbiased way. {\tt CPROPS} first assigns the measured emission to clouds by identifying emission above a $4\sigma$ noise level and then decomposing the emission into individual clouds \citep[see][for further details]{Rosolowsky:2006}. It then extrapolates physical properties such as cloud sizes and masses to 0K noise level, independent of the beamsize (i.e. deconvolved).  Using {\tt CPROPS}, we find 11 clouds all of which have also been identified as such by eye. The cloud properties as identified by {\tt CPROPS} are presented in the lower part of Table \ref{tab.clouds}. We were not able to setup {\tt CPROPS} in such a way that all the eye-identified clouds were found without a large number of clouds we do not believe are real being also identified by {\tt CPROPS}.
{\tt CPROPS} was used with the following parameters: a constant distance {\tt DIST} equal to $490$ kpc; the {\tt /NONUNIFORM} parameter along with a custom noise map computed from velocity channels without any signal (i.e. outside the rotation curve), to take into account non uniform noise over the CO map; the following values for the decomposition parameters {\tt FSCALE=2.0, SIGDISCONT=0} to ensure that each area of non-contiguous emission is assigned to an individual independent cloud.

Using the identified cloud emissions, {\tt CPROPS} computes four initial quantities through an extrapolation to a $0\ $K noise level: the $\sigma_{x}$ and $\sigma_{y}$ spatial dispersions, the velocity dispersion $\sigma_{v}$ and the CO flux ${\rm F_{CO}}$. From these quantities, the following physical quantities are deduced. An effective radius $R_{e}$ (col. 6) is obtained using the formula:
\begin{eqnarray}
R_{e}& = &\eta \sigma_{r}\\
\label{eq.Re}
\sigma_{r}& = &\sqrt{\sqrt{\sigma_{min}^2-\left(\frac{FWHM_{beam}}{2\sqrt{2\ln2}}\right)^2}\sqrt{\sigma_{max}^2-\left(\frac{FWHM_{beam}}{2\sqrt{2\ln2}}\right)^2}}
\end{eqnarray}
Where $\eta=3.4/\sqrt \pi=1.92$ is an empirical geometric factor from \citet{Solomon:1987} to take into account the radial distribution of the gas density inside the molecular cloud and $\sigma_{min}$ and $\sigma_{max}$ are spatial dispersions along the major and minor axes of the cloud.

 For a few of the clouds identified by {\tt CPROPS}, the minimum dispersion $\sigma_{min}$ was found to be smaller than the beamsize. In these cases we have chosen to use an arbitrary value of $\sigma_{min}$ corresponding to an effective minimum spatial dispersion of $10$pc. 

We have also computed an extrapolated radius (col. 7)
\begin{equation}
R_{ex}=\frac{\sqrt{A_{2\sigma}-(FWHM_{beam}/2\sqrt{2\ln2})^2}}{\pi}
\end{equation}
where $A_{2\sigma}$ is the area of the individual clouds extrapolated down to a $2\sigma$ level projected on the sky plane \citep[see][for details]{Rosolowsky:2006}.

The clouds' integrated CO(2--1) intensities (col. 8, lower part of Table \ref{tab.clouds}) were obtained for each cloud by dividing the {\tt CPROPS} computed ${\rm L_{CO}}$ luminosity  by the projected area of each individual cloud, $A_{2\sigma}$, at the two sigma level.

\subsubsection{Comparing with eye identification}

For the subset of clouds which have both been identified by eye and by the {\tt CPROPS} package, we can compare the physical properties obtained by the two independent methods. Table \ref{tab.clouds} shows the properties computed by hand for the 15 clouds identified by eye, and by {\tt CPROPS} for the first eleven which have also been identified with {\tt CPROPS}. Each property is computed slightly differently for the two methods and the next paragraphs will explain these differences.

In the manual identification method, the offsets for the cloud positions are obtained by taking the average position in right ascension and declination of the pixels inside of the half maximum contour of the cloud emission. In the case of {\tt CPROPS}, the offsets are equal to the first moments of the cloud emission (down to $2\sigma$) along the right ascension and declination axis. The positions are in general the same to within 0.2 beam fwhm.

The systemic velocity is taken as the average of the gaussian fit to the CO line (see Sect. \ref{sec.cloudID}) for the manual identification method, in the case of {\tt CPROPS} it is computed as the first moment along the velocity axis of the cloud emission down to $2\sigma$. Following the same idea, the line width are computed in the manual case as the the function width at half maximum of the fitted gaussian and as the second moment converted to FWHM by multiplying by $2\sqrt{2\ln2}$ for the properties derived from {\tt CPROPS}. An effective radius $R_{e}$ is obtained using Eq. \ref{eq.Re} and preceding section.
%Redondant avec partie prŽcedente
%\begin{eqnarray*}
%R_{e}^{\tt CPROPS}& = &\eta\ \sigma_{r}\\
%\sigma_{r}& = &\sqrt{\sqrt{\sigma_{min}^2-\left(\frac{FWHM_{beam}}{2\sqrt{2\ln2}}\right)^2}\sqrt{\sigma_{max}^2-\left(\frac{FWHM_{beam}}{2\sqrt{2\ln2}}\right)^2}}\\
%\end{eqnarray*}
%Where $\eta=3.4/\sqrt \pi=1.92$ is an empirical geometric factor from \citet{Solomon:1987} taking into account the radial distribution of the gas density inside the molecular cloud. For a few of the clouds identified by {\tt CPROPS}, the either the minimum (clouds ) or both (clouds ) of the computed deconvolved dispersions where found to be smaller than the beam size. In these cases we have chosen to use an arbitrary value of spatial dispersions corresponding to an effective radius of $10$pc. 
In the case of the manual identification, the effective radius was obtained in the following way.
\begin{equation}
R_{e}^{eye} =  \frac{1.92}{2\sqrt{2\ln2}}\sqrt{A_{\frac{1}{2}}-A_{beam}}
\end{equation}
where $A_{\frac{1}{2}}$ is the area in ${\rm pc^2}$ inside the contour at half of the peak integrated intensity for each cloud. And $A_{beam}=1270 pc^2$ is the area subtended by the $15\arcsec$ beam at a distance of $490kpc$. The $2\sqrt{2\ln2}$ factor converts the FWHM to dispersion and the $1.92$ from dispersion to effective radius.
%We have also computed in the case of {\tt CPROPS} an extrapolated radius:
%\begin{equation}
%R_{ex}=\frac{\sqrt{A_{2_{\sigma}}-(FWHM_{beam}/2\sqrt{2\ln2})^2}}{\pi}
%\end{equation}
%where $A_{2\sigma}$ is the area of the individual clouds down to a $2\sigma$ level projected on the sky plane.

The CO intensity was computed using {\tt CPROPS} CO luminosity and dividing it by the area $A_{2\sigma}$ of the cloud computed above. In the case of the manual selection, the emission inside the 50\% level contour was summed over the velocity range $V_{o}\pm \Delta V$ and multiplied by the $0.41 \kms$ channel width.

In both cases, the virial mass was obtained using the following formula from \citep{Solomon:1987}: 
\begin{equation}
M_{vir} = 1040R_{e}\left(\frac{\Delta V}{2\sqrt{2\ln2}}\right)^2
\end{equation}
The uncertainties in the virial mass estimates are dominated by the hypothesis that the molecular clouds are indeed gravitationally bound and by the value of the geometric factor describing the density distribution of the gas. {  The marginally gravitationally bound case of a cloud in isolation with no magnetic field would yield masses a factor 2 less than virial. The virial masses are widely used because clouds have magnetic fields, are not isolated, and collapse to form stars.}

The ${\rm H_{2}}$ masses (col. 9 for top and col. 10 for bottom parts of Table \ref{tab.clouds}) are derived following the method described in Sect. \ref{sec.mass8mu}, using the 50\% contour in the case of the manual identification and the $2\sigma$ contour for {\tt CPROPS}.

The last column in Tab \ref{tab.clouds}, $\ratioo$ is the ${\rm H_{2}}$ column density derived in Sect. \ref{sec.mass8mu} divided by ${\rm I_{CO}}$.

\subsection{Estimation of errors}
For the CPROPS identification method, the uncertainties for the quantities FWHM, ${\rm R_{e}}$, ${\rm I_{CO}}$ and ${\rm M_{vir}}$ are computed using the bootstrapping method of CPROPS \citep[see details in ][]{Rosolowsky:2006}. We now describe the computation of the uncertainties in the case of the identification of clouds by eye. The FWHM line width and associated error are computed with the CLASS software gaussian line fitting algorithm. Then, using the line widths obtained, the unvertainty in integrated intensity $\sigma_{Kkm/s}$  is calculated for each cloud.   The uncertainty on the size ${\rm R_{e}}$ is estimated by calculating ${\rm R_{e}}$ from contours (cloud sizes) placed at $I_{peak}/2 - \sigma_{Kkm/s}$ and $I_{peak}/2 + \sigma_{Kkm/s}$, thus bracketing the cloud size obtained using the $I_{peak}/2$ contour.  This gives respectively an upper (lower) bound on the value of ${\rm R_{e}}$. The errors are then propagated into the Virial mass. For each cloud, the contours defined at $I_{peak}/2 \pm \sigma_{Kkm/s}$ are used to compute the molecular gas mass (see Sect. 6) from the 8 micron map and thus estimate the uncertainties in the molecular gas mass (Tab. 2 col. 9) and the $\ratioo$ factor (Tab. 2 col. 10).

\section{The size-linewidth relation for the molecular clouds in NGC 6822}

\begin{figure}[tp]
\begin{flushleft}
\includegraphics[angle=270,width=88mm]{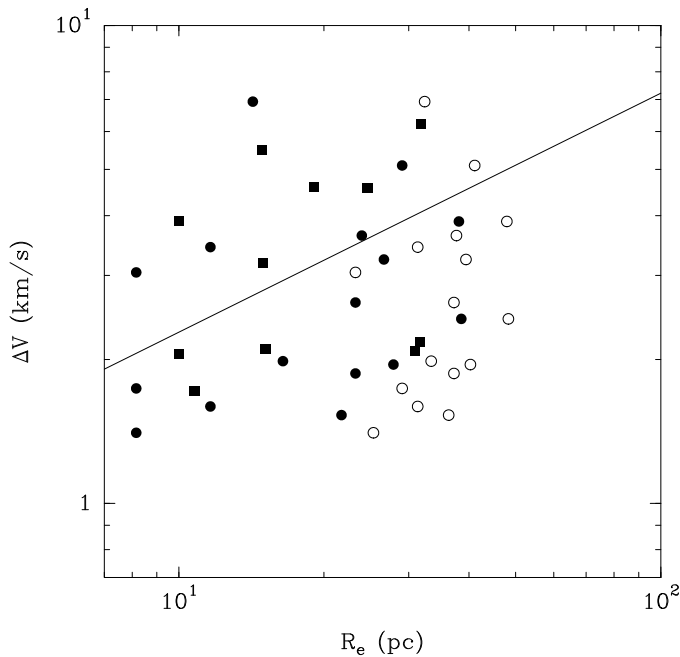}
\caption{Size vs linewidth for molecular clouds in NGC~6822, open and filled circles correspond respectively to measured and deconvolved sizes (see Sect. \ref{sec.clouds}). Filled squares correspond to physical properties computed with the {\tt CPROPS} package. ${\rm \left.\Delta V\propto {  R_{e}}\right.^{0.5}}$ is the galactic relationship from \citet{Solomon:1987}. Notwithstanding considerable
scatter, the distribution of the clouds (corrected for finite
beamsize) in this diagram appears consistent with a size-linewidth
relation similar to that in the Milky Way.}
\label{fig.size_width}
\end{flushleft}
\end{figure}

\begin{figure}[tp]
\begin{flushleft}
\includegraphics[angle=270,width=88mm]{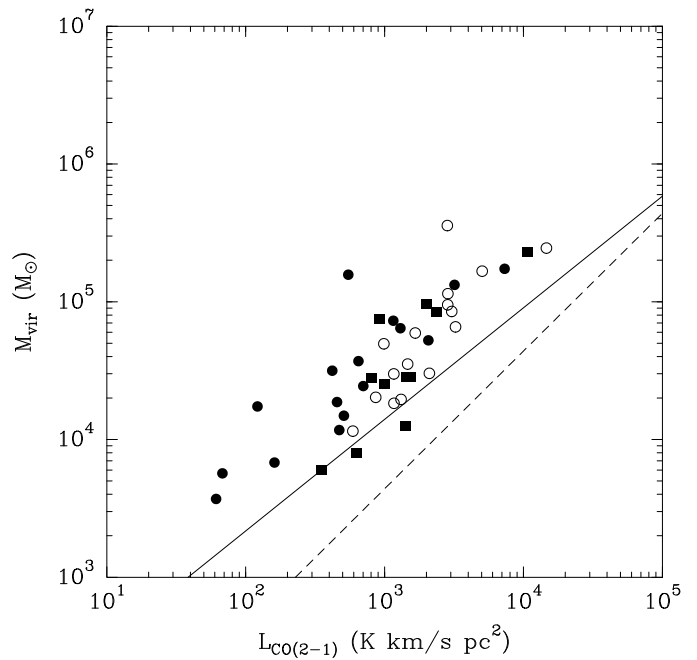}
\caption{Virial mass vs CO luminosity for molecular clouds in NGC~6822, open and filled circles correspond respectively to measured and deconvolved sizes (see Sect. \ref{sec.clouds}). Filled squares correspond to physical properties computed with the {\tt CPROPS} package. The line ${\rm M_{vir} = 9796 \left(\frac{L_{CO(1-0)}}{1000} \right)^{0.8}}$ is the galactic relationship from \citet{Solomon:1987}, the dashed line corresponds to a constant galactic $\ratioo$ ratio of $2\times10^{20} \Xunit$. The virial masses are systematically larger for a given luminosity in NGC~6822 compared to clouds in the Milky Way.}
\label{fig.mvir}
\end{flushleft}
\end{figure}

Figures \ref{fig.size_width} and \ref{fig.mvir} show respectively the size-linewidth relation ($\Delta V$ vs. $R_{e}$) and the virial mass vs
CO luminosity, showing in both cases the Galactic values taken from \citet{Solomon:1987} as a straight line. The distribution of the clouds in NGC~6822 appears consistent with a size-linewidth
relation similar to that in the Milky Way GMCs.
In recent work \citet{Heyer:2009} obtain lower H$_{2}$ masses and a dependency on the square root of the surface density, the variation we obtain in these parameter is much to low to reach a conclusion. Like \citet{Solomon:1987}, they concluded that GMCs are gravitationally bound.

Fig. \ref{fig.mvir} shows that there is a factor several difference between the virial 
masses of the NGC~6822 clouds and the masses obtained from the 
size and CO(2--1) intensities using a Galactic $\ratioo$ conversion factor.  
For a ${\rm {CO(2-1)}/{CO(1-0)}}$ ratio of 0.7, the difference is $5\mbox{ -- }6$, the true $\ratioo$ value is then at least $5\mbox{ -- }6$ times higher than in the Milky Way.  Due to the low metallicity of NGC~6822, cloud sizes as seen in
CO are probably underestimated compared to Galactic observations, where the
shielding will be much more efficient to protect CO molecules and allow 
the CO size to be similar to the total size of the H$_2$ dominated region (the molecular cloud).
Because the line width is presumably determined by the total mass, 
%but the size seen in CO observations is limited to roughly the region which is optically thick in CO, 
the virial masses should be underestimated
linearly with the size. CO luminosities, used when applying a
$\ratioo$ factor, should be underestimated twice as much in proportion
because luminosities vary with the square of the size.
In the following section we try to estimate the $\ratio$ factor by means without these drawbacks.

 \section{The total molecular mass of NGC~6822}
 \label{sec.mass}
 In this section we describe how we make a CO integrated intensity map to trace the H$_2$ column density. In order to test whether there could be substantial amounts of molecular gas without associated CO emission, we use two other alternative methods \citep[similar to][]{Israel:1997} to estimate H$_2$ masses for comparison.
\subsection{CO(2-1) intensity maps}
\begin{figure*}[tp]
\begin{flushleft}
\includegraphics[angle=270,width=180mm]{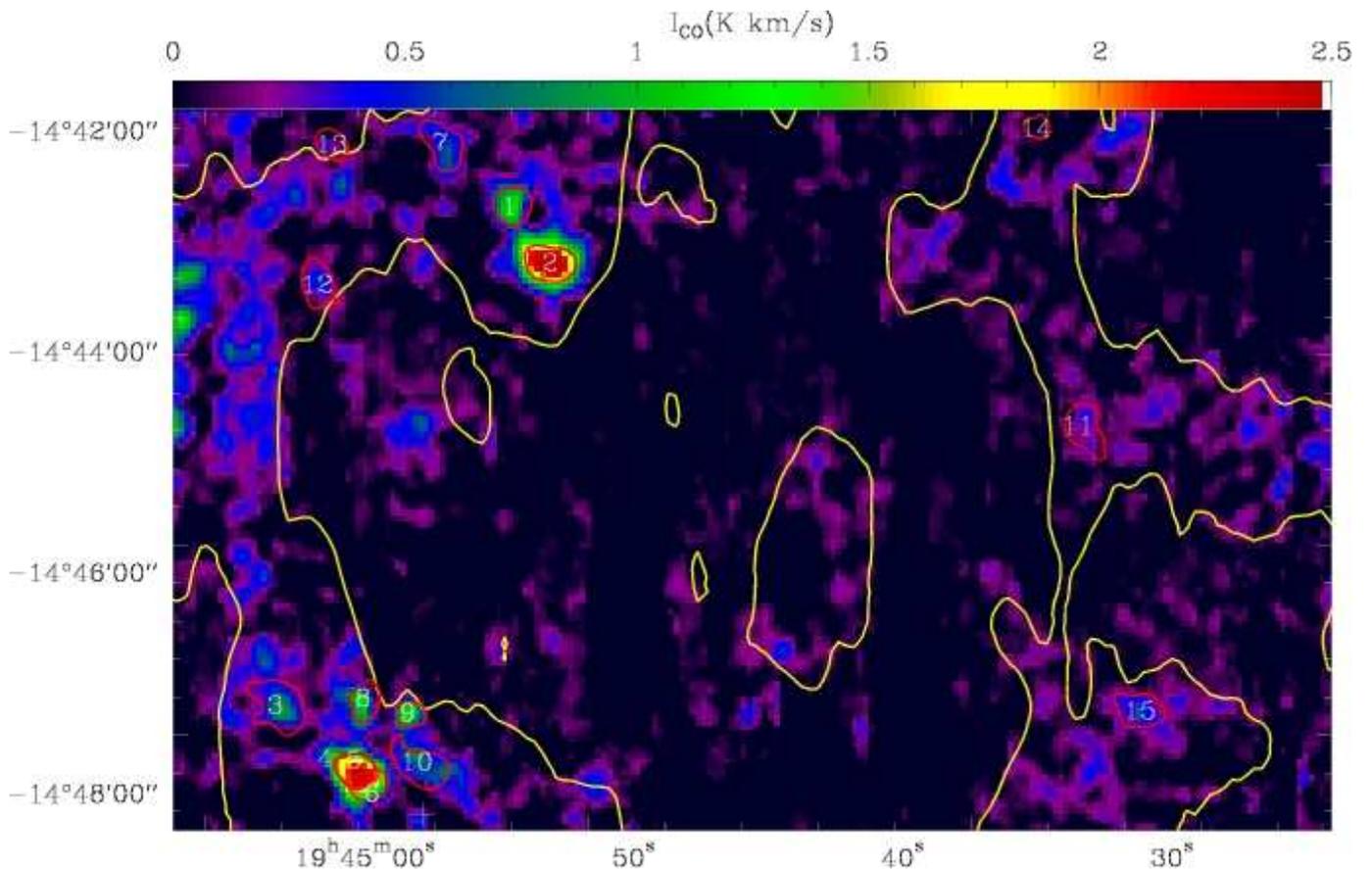}
\caption{CO(2-1) intensity map obtained using the masking method described in Sect. \ref{sec.mask}. The map unit is K$\kms$, the yellow \ion{H}{I} contour is at $\NHI=10^{21}\ {\rm cm^{-2}}$, the red contours are the half maximum levels of each cloud identified by eye, the numbers correspond to the ones in Table \ref{tab.clouds}. The cross corresponds to (0,0) offsets of the HERA CO(2-1) map.}
\label{fig.COmom0}
\end{flushleft}
\end{figure*}
\label{sec.mask}
We compute the CO(2-1) integrated intensity map using a masking method, taking into account the 21cm atomic hydrogen line data, we developed in order to filter out some of the noise present in the observations and increase the sensitivity to low intensity possibly diffuse CO emission. Previous masking methods used masks created from spatially smoothed versions of the original CO data cubes to filter out regions dominated by noise \citep{Adler:1992,Digel:1996,Loinard:1999}.

We use the 21cm atomic hydrogen data at $12\arcsec\times42\arcsec \times 1.6\kms$ resolution \citep{de-Blok:2006} to achieve the same goal, the underlying hypothesis being that molecular gas is unlikely to be present for low enough values of $\NHI$ so the corresponding velocity channels can be discarded when computing the integrated intensity CO map.  For each pixel of the \ion{H}{I} cube, we estimate a noise level from velocity channels that clearly contain no signal from NGC 6822, we then create a binary mask keeping only the velocity range for each pixel corresponding to a \ion{H}{I} signal value above a defined factor of the pixel noise. Since the noise in the \ion{H}{I} cube varies little over the region observed in CO, a cut in S/N is like a cut in antenna temperature. The integrated moment map for the CO(2-1) data (Fig. \ref{fig.COmom0}) is then computed summing only velocity channels included in the \ion{H}{I} mask. 
The result is an increased S/N ratio as the channels contributing only noise to the sum are no longer taken into account.
The value of the noise threshold was chosen at $6\sigma$ which corresponds a map averaged \ion{H}{I} brightness temperature of 30K.
We tested masking values between 25 and 40~K (5 to 8 $\sigma$) and the total CO intensity varied by only a few percent. Significantly above or below these values, CO signal is lost or more noise is included.
Using this procedure we miss the very weak cloud 13 and part of cloud 14 shown in Fig. \ref{fig.spectra}.

The values in the CO integrated intensity map (Fig. \ref{fig.COmom0}) yield H$_2$ column densities
when multiplied by a $\ratioo$ factor.  If we sum all of the emission in Fig. \ref{fig.COmom0}, we obtain
about L$_{\rm CO(2-1)} \sim 5.5 \times 10^4$ K km s$^{-1}$ pc$^2$, or some $3 \times 10^5$ M$_{\sun}$ 
in the region we have observed for a $\ratioo$ ratio of $2.0\times10^{20}\ \Xunit$. Our HERA map covers an area corresponding to 40\% of the total \emph{Spitzer} ${\rm 70\mu m}$ luminosity of NGC 6822 but over 60\% at the 1Mjy/sr cutoff we apply later to be less affected by the noise. Thus for all of the galaxy we can expect the total CO(2--1) luminosity to be between 1.5 and 2.5 times our value. Assuming a ratio of 0.7 between {CO(2--1)} and {CO(1--0)} we can estimate a total {CO(1--0)} luminosity of $16\pm4\times10^4\kms\ {\rm pc^2}$ in substantial agreement with the value $12(+12,-6)\times10^4 \kms\  {\rm pc^2}$ \citet{Israel:1997} estimates for the whole galaxy.
%CHECK exact values
The CO emission is thus rather weak and the next step is to compare with the other means of locating molecular gas.

\subsection{Infrared data} 
All of the infrared maps are taken from the SINGS (Spitzer Infrared Nearby Galaxies Survey, \citealt{Kennicutt:2003}) fifth public data delivery.
The 8\mum  map does not show a morphological similarity with the local
galactic emission as seen in CO; we have therefore neglected the Milky Way cirrus contribution in this band.
For the 160\mum , 70\mum and 24\mum MIPS band we have used the maps from \citep{Cannon:2006} where a  
smooth component fitted on emission outside NGC~6822 representing  the  
local emission and instrumental and observational bias has  
been substracted.
The IRAS 100\mum data was obtained using the IPAC HiRES algorithm using default parameters with no additional smooth component subtracted.

\subsection{Molecular gas mass from 8${\rm \mu m}$ emission}
\label{sec.mass8mu}

\begin{figure}[tbp]
\begin{flushleft}
\includegraphics[angle=0, bb= 44 310 306 580 , width=88mm, clip]{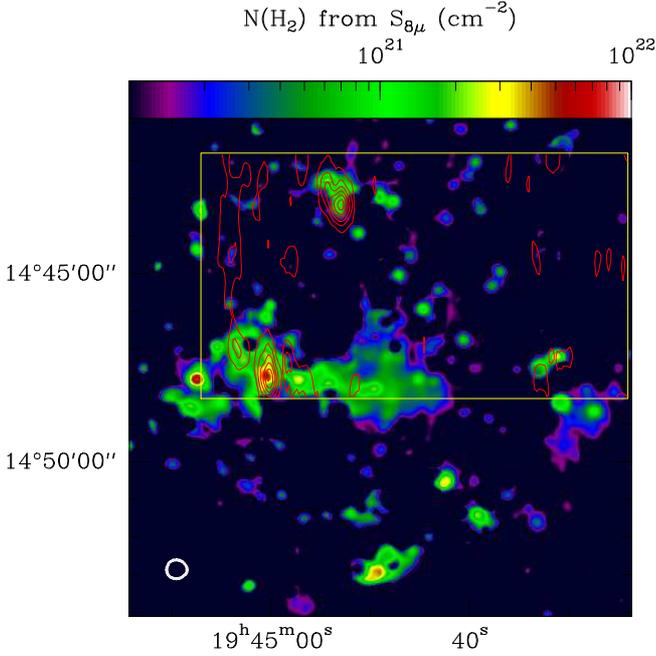}
\caption{
H$_{2}$ column density map estimated from PAH ${\rm 8\mu m}$ emission, CO(2-1) intensity contours every $(0.15\mathrm{K}\kms)$ starting from $(0.15\mathrm{K}\kms)$. The yellow rectangle is the region we have mapped in CO(2--1) with HERA. The beams  for the NH$_{2}$, CO are plotted in white.}
\label{fig.NH2_8mu}
\end{flushleft}
\end{figure}

%\st{The emission of the large PolyAromatic Hydrocarbon (PAH) molecules 
%present in the interstellar medium is a means of locating neutral gas.}
Before attempting to use the 8 \mum emission to trace gas, the ${\rm 8\mu m}$ data was corrected for stellar continuum emission by subtracting the $3.6\mu m$ emission scaled by a factor 0.232, following \citet{Helou:2004}. In the 8\mum PAH band, we expect that  less than 10\% of the emission is from hot dust. This comes from extrapolating a blackbody curve from the 24\mum point in Fig. 12 from \citet[][assuming the dust emitting at 24\mum can be considered ``hot'']{Draine:2007} to 8\mum and comparing that to the 8\mum emission on the curve.  At most, this would slightly reduce the gas masses we calculate for Hub V and X.  We have therefore not subtracted a hot dust contribution from the 8\mum PAH emission.

PAHs have often been used as tracers of Star Formation, including in high redshift objects for which very little is known \citep[e.g.][]{Aussel:1999}.
In this section we propose to use the PAH emission observed in band IRAC4 (\emph{Spitzer}) to trace the gas.
There is a close theoretical relationship between the PAH band emission per H-atom (atomic or molecular) and the incident UV field up to UV fields of 10000 times solar \citep[][Fig. 13, lower panel]{Draine:2007}  The UV fields in NGC~6822 are far below this value.  In cloud cores, where UV emission is not available to excite the PAHs, very little 8$\mu$m emission is expected.  However, GMCs are quite porous to UV radiation \citep{Boisse:1990} so we expect to see a rather thick cloud surface, made thicker in a low metallicity object like NGC 6822.  Cloud cores make up only a small fraction of the molecular mass of a galaxy and this is particularly clear for NGC 6822  from the  weak \, \inst{13}CO and HCN emission (Sect. \ref{sec.HubV}).  As shown in Sect. \ref{sec.clouds}, we see individual GMCs in NGC~6822 (individual because the narrow lines cannot come from an accumulation of objects) and the majority of them are of order our beam size, i.e. not unresolved, and similar in size to Galactic GMCs.  \citet{Bendo:2008} show that the PAH emission at large (kpc) scales in nearby spirals seems to trace the cool diffuse dust responsable for most of the 160$\mu$m emission, thereby tracing the gas mass. \citet{Regan:2006} conclude from their observations that the PAH emission at large scales can be used to trace the interstellar medium.  Thus, from both an observational and theoretical point of view, {\it at large scales} the PAH emission can be used to trace neutral gas.
The lower metallicity in NGC 6822 is not an issue for our method because we use regions with little or no star formation and low 70 and 160 micron emission, such that little or no molecular gas is expected, in order to ``calibrate" the 8$\mu$m emission per H-atom per FUV ratio.

We compute the emissivity of the PAHs per hydrogen atom and per unit of ISRF (traced by the GALEX FUV data) in the 
${\rm 8\mu m}$ \emph{IRAC} band, $((S_{8\mu m }/FUV)/\NHI)_{o}$, in regions 
far from major star forming regions and with low but well-measured \ion{H}{I} 
column densities.  In these regions we find $(\NHI/S_{8\mu m })_{o}=1.25\times10^{22}$ H cm$^{-2}$ MJy$^{-1}$ sr, close to the value $1.1\times10^{-26}$ erg s$^{-1}$ sr$^{-1}$ H$^{-1}$ given by \citet{Draine:2007} in their Table 4.
 
%%Calcul
% NHI/8mu = 1.25 10^22 H cm^-2 MJy-1 sr
% 8/NHI = 8e-17Jy/sr cm2/H = 8e-17 1e-27 1e7 1e-4 = 11.6e-27 erg/s/sr/H

The value of the interstellar radiation field at each position was derived from the far ultraviolet GALEX map and the mean value of the ISRF given by \citep{Draine:2007a} for the whole galaxy.   No corrections were made to correct for UV extinction by Milky Way dust. The GALEX FUV data were taken from the GR5 public release of the MAST archive

%For the following, we 
%make the uncertain assumption that PAHs have similar emissivities in
%atomic and molecular media.   However, the PAH emission appears to be a good 
%tracer of the radiation field  and the neutral H column density
%over a wide range of intensities and 
%apparently irrespective of whether the gas is atomic or molecular
%as long as it is neutral \citep{Draine:2007}.

% Fig 13 bottom panel -- cross-sect unvarying per H and per unit rad intensity
%In such regions where there 
%is no molecular gas, $(S_{8\mu m }/\NHI)_{o}$ equals $S_{8\mu m }/\NHI$. 
Then, from the ${\rm 8\mu m}$ per H-atom emissivity, the ${\rm 8\mu m}$ emission, FUV emission and the 
\ion{H}{I} column density at each position, the molecular gas column density can be 
estimated at each position as follows:
%We use this emissivity in the absence of molecular gas to estimate the 
%molecular gas column density using both the atomic hydrogen column 
%density and the PAH ${\rm 8\mu m}$ emission
\begin{equation}
2\  \NHmol= \left(\frac{\NHI}{S_{8\mu m}/FUV}\right)_{o}S_{8\mu m}/FUV - \NHI
\end{equation}

{  At the 15$\arcsec$ resolution used for 8\mum and FUV, the atomic line data is assumed to be smooth thus the original $12\arcsec \times
42\arcsec$ HI map is directly subtracted. 
}
\citet{Draine:2007} find that the emissivity per H-atom and per unit of interstellar radiation field is constant over a wide range of values of the ISRF and that this should be true irrespective of whether the gas is atomic or molecular as long as it is neutral.

%This method implicitly assumes that the radiation field, linked to the per H-atom emission emissivity, is constant.
Taking into account the variation of the PAH emissivity with the radiation field does not (significantly) change the computed masses of individual clouds except in the case of Hub V (cloud 2 in Table \ref{tab.clouds}) where the mass is found to be 5 times smaller than for the non ISRF corrected case, for an estimated interstellar radiation field of 50 Habing.

In extreme radiation fields, PAHs can be destroyed.  However, in NGC 6822 very little of the dust mass is exposed to such fields; according to Draine et al. 2007 Table 5, less than 1\% of the dust in NGC 6822 is exposed to a high ISRF.  Furthermore, there is no evidence of PAH destruction through low 8/24 \mum ratios (cf. Table 2 of Cannon et al. 2006).  PAH destruction is unlikely to affect our estimate of the H$_{2}$ mass.

The spatial correlation between the 8 \mum PAH emission and the FUV is quite good.  The presence of zones with an FUV peak but without a PAH peak does not affect our mass estimates and the opposite, PAH emission adjacent FUV emission coming from far enough away that the division would not affect the same pixels, would only cause us to overestimate the H$_{2}$ mass.

The corresponding H$_2$ column density map is shown in the upper panel 
of Fig. \ref{fig.NH2_8mu}, which shows the H$_2$ distribution in NGC~6822 
derived as above along with the CO(2--1) integrated intensity contours for the
part observed by us in CO.  

This way, we estimate the molecular gas mass within our observed zone
to be ${\rm M(H_2)\sim 6\times10^6\ M_{\sun}}$ and about 
${\rm M(H_2)\sim1\times 10^7\ M_{\sun}}$ over NGC~6822 as shown in Fig. \ref{fig.NH2_8mu}.
%routine nh2_8mu.greg
This is 20 -- 25 times greater than using a Galactic value and is in excellent agreement
with \citet{Israel:1997} who estimated the total H$_2$ mass to be about $1.5 \times 10^7$ M$_{\sun}$. 

We have also calculated the H$_2$ masses for the individual clouds in Table 2 
using the 8${\rm \mu m}$  emission and the last column gives the corresponding estimated $\ratiot$ values.
The median and average values are 15 -- 20 times the Galactic value, assuming 
a line ratio of 0.7 to go from $\ratiot$ to $\ratio$. 
% {\sc NOT SO TRUE ANYMORE This suggests that the virial masses 
%underestimate the cloud masses by a factor 1.5 -- 2, presumably because the clouds
%are smaller in CO than their real H$_2$ size.}
%Since the 8${\rm \mu m}$  was measured where there is little star formation, 
%the value of $\NHI/S_{8\mu m}$ will be overestimated in regions with 
%active star formation, tending to overestimate the amount of H$_2$ near
%\ion{H}{I}I regions.  No CO emission is detected directly from Hubble X, although
%there is a peak in the 8${\rm \mu m}$  based H$_2$ column density map 
%suggesting that there should be $ 1- 3 \times 10^5$ M$_{\sun}$ of H$_2$.
%However, this is much less than in Hubble V, which has a comparable
%UV and H$\alpha$ luminosity.  Only a very slight increase in the 8${\rm \mu m}$ 
%emissivity due to a strong radiation field is necessary to create an overestimate
%of this order in the H$_2$ masses of large star-forming regions.
Overall, the agreement between the CO emission and the 8${\rm \mu m}$  based
vision of where the molecular gas is found is quite good, suggesting that CO 
traces the H$_2$, albeit with a much higher $\ratio$ factor than
in the Galaxy. Galactic CO emission is present towards NGC~6822 \citep{Israel:1997} (see Sect.
\ref{sec.cirrus}), the \ion{H}{I} column is of course unaffected but the 8${\rm \mu m}$  continuum
could be. However, since the morphology of the H$_2$ column density map resembles NGC~6822 so closely but not the local emission, we
expect this contribution to be low.

%\citet{Draine:2007} have shown that 
%$$u_{\nu}=Uu_{\nu}^{MMP83}$$
%For values of the ISRF in units of the ISRF estimated by \citet{Mathis:1983} for the solar neighborhood. The PAH emission

\subsection{Molecular gas mass from 160${\rm \mu m}$ emission}
\label{sec.160mu}
\begin{figure*}[htbp]
\begin{flushleft}
\includegraphics[angle=0, bb= 44 53 494 310 , width=170mm, clip]{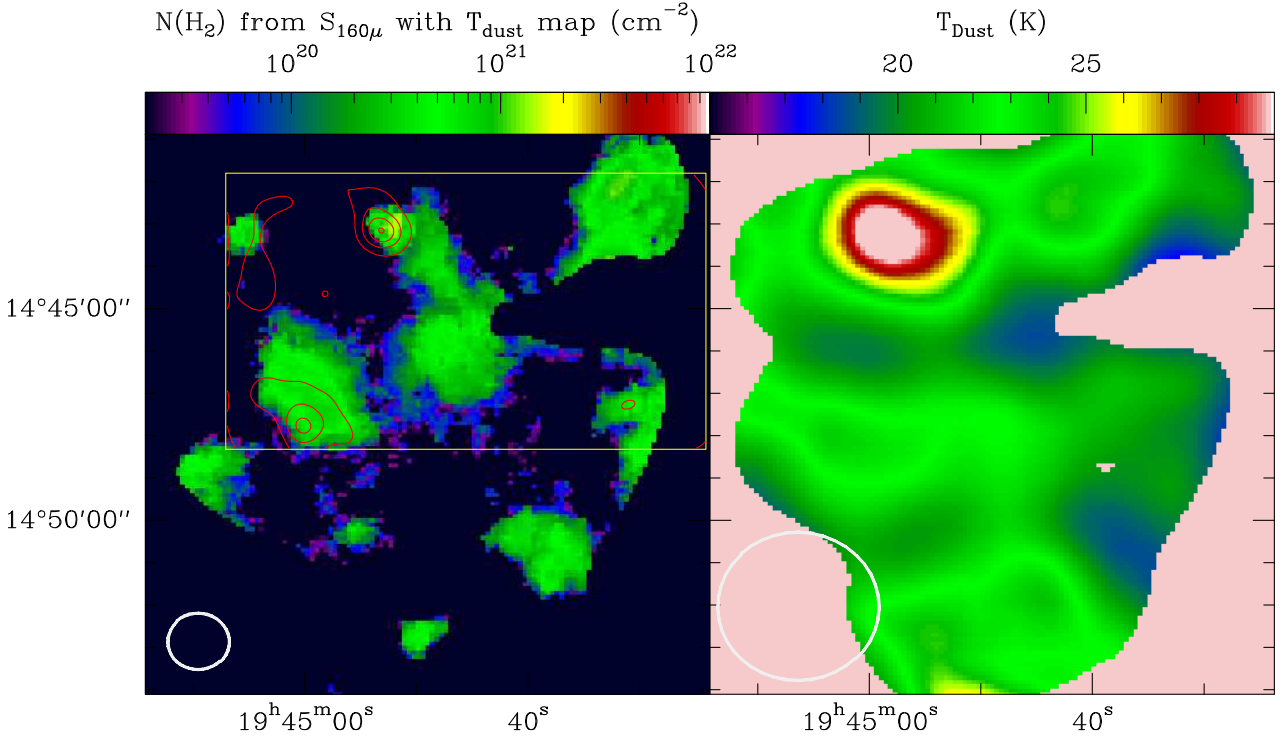}
\caption{
H$_{2}$ column density map estimated from $160{\rm \mu m}$ emission using the dust temperature map shown on the right (\emph{left}), CO(2-1) intensity contours every $(0.15\mathrm{K}\kms)$ starting from $(0.15\mathrm{K}\kms)$. Temperature map from $160{\rm \mu m}$ Spitzer and $100{\rm \mu m}$ IRAS data (\emph{right}) . The yellow rectangle is the region we have mapped in CO(2--1) with HERA. The beams  for the NH$_{2}$, CO and temperature} are plotted in white.
\label{fig.NH2_160mu}
\end{flushleft}
\end{figure*}

We can perform the same sort of calculation based on the 160${\rm \mu m}$  emission, {  which we assume to come from  dust grains large enough that they reach thermal equilibrium and are not affected by individual photons. The properties of the grains can be found in \citet{Li:2001} and \citet{Draine:2007}.}
%\st{The first step is to measure or assume a 160${\rm \mu m}$  dust cross-section.
%The cross-section scales with the metallicity so we take a value of
%$0.3 \times \sigma_{\rm DL84}$ where 
%$\sigma_{\rm DL84} = 2.7 \times 10^{-25}$ cm$^2$ is the cross-section derived
%by Draine:1984 at 160${\rm \mu m}$. This value is very close to the average value measured
%in the regions assumed to be without H$_2$ that were used to estimate $\NHI$ from ${\rm 8\mu m}$ emission.}
 The first step is to measure the dust temperature.
Since the work by \citet{Israel:1997}, the Spitzer data for NGC~6822 has become 
available \citep{Cannon:2006}, extending to longer wavelengths than IRAS and thus 
more sensitive to cool dust.  Using the 70${\rm \mu m}$  and 160${\rm \mu m}$  Spitzer data, we 
derive, like \citet{Cannon:2006}, dust temperatures around 25~K (assuming that the dust cross-section varies with 
$\lambda^{-2}$).  Since cool dust emits very weakly at 70 \mum, we chose to use the 160${\rm \mu m}$ Spitzer and 100 ${\rm \mu m}$ IRAS data to better measure the temperature of the cool dust component.
However, some of the 100${\rm \mu m}$  emission may still come from a warm component, causing an 
overestimate of the dust temperature of the cool component and a corresponding underestimate of the 
gas mass.  NGC~6822 has an SED \citep[][figure 12]{Cannon:2006} similar to that of NGC 4414, as measured by the ISO LWS01 scan by \citet{Braine:1999}.  In their Figure 4, they present a breakdown of the dust emission into warm and cool components.  The 100 $\mu$m emission due to the warm dust is about 8\% of the total 70 micron emission.  In equation \ref{eq.XX}, we therefore take $\chi = 0.08$ as our fiducial value but also test $\chi = 0$ and $\chi = 0.16$ to measure the effect of an error in $\chi$.  We smoothed the 160${\rm \mu m}$  data to the resolution of the IRAS 100 \mum 
HiRes maps ($118\arcsec \times 102\arcsec$).  Assuming a modified grey-body law with a spectral index $\beta=2$ for the dust, we then estimate dust temperatures around 21K with Eq. \ref{eq.YY} instead of 23K using the 70 and 160 \mum data for the same regions.  Changing $\chi$ from 0 to 0.08 causes a 0.3K change in dust temperature.

\begin{equation}
\label{eq.XX}
S_{100\mu m_{cool}}=S_{100\mu m}-\chi S_{70\mu m }
\end{equation}

\begin{equation}
\label{eq.YY}
T_{dust}=f\left(\frac{S_{100\mu m_{cool}}}{S_{160\mu m}}\right)
\end{equation}

The computed dust temperature map is shown at the right of Figure \ref{fig.NH2_160mu}.  The temperature map does not cover the whole galaxy because it was necessary to clip the very low signal-to-noise regions.  Cuts were applied at 4 MJy/sr for both the 100 and 160 \mum maps, leaving the odd shape seen in Figure \ref{fig.NH2_160mu}.
From the dust temperature, the HI column density, and the 160 \mum  emission, assumed optically thin, we can estimate the dust cross-section at 160 \mum which can then be applied to regions with molecular gas.

%\st{To test this, we smoothed the 160${\rm \mu m}$  data to the resolution of the IRAS
%HiRes maps ($118\arcsec \times 102\arcsec$) and calculated the temperature from the 100${\rm \mu m}$/160${\rm \mu m}$ 
%flux ratio. 
%This consistently yielded a lower temperature, as can be seen in the bottom right panel of Fig. fig.NH24panels}.

Although the map of dust temperature is at IRAS resolution, we apply it
to the full resolution 160 \mum data so that the morphology is better reproduced.  Smoothing does not affect the total 160 \mum flux.
Averaging the HI/160mu ratio map over regions with high enough 160 emission that the noise has little effect (4 MJy/sr)
and low enough 160 emission to exclude regions where molecular gas will be present (8 MJy/sr), we obtain dust cross-section at 160 \mum of $\sigma_{160} = 5.1 \times 10^{-26}$cm$^2$ per H-atom.  We also varied the threshold values (from 4--8 to 6--12), this lead to a variation of sigma of at most 10\%.

\citet{Cannon:2006} found consistent values of Mdust/MHI for the individual regions they observed and some of the scatter is certainly attribuable to the molecular gas that they could not measure, although in most cases the HI dominates.  They found a factor ~5 lower ratio when they calculated Mdust/MHI for the whole galaxy.  Using the 70/160 dust temperature like Cannon and the 160 micron emission, but only over the area where we felt we could reliably estimate the dust temperature (cuts at 1 MJy/sr at 70 \mum and 4 MJy/sr at 160 \mum, both smoothed to the 160 \mum resolution)
we find a dust cross-section (equivalent to Mdust/MHI ratio) of $4.7\times10^{-26}$ cm$^2$ per H-atom with little variation, unlike \citet{Cannon:2006}.

Assuming a linear dependence of $\sigma_{160}$ on metallicity (Oxygen abundance), a solar oxygen abundance of $12+log(O/H) = 8.67$ \citep{Asplund:2005},
$12+log(O/H) = 8.11$ in NGC 6822 \citep{Lee:2006}, and a solar metallicity dust cross-section of $\sigma_{160} = 2.25 \times 10^{-25}$cm$^2$ per H-atom \citep[Table 6 of ][]{Li:2001}, we obtain $\sigma_{160} = 6.3 \times 10^{-26}$cm$^2$ per H-atom for NGC 6822.
Thus, our ``observational'' results are in good agreement with model calculations.  

Using this temperature map and the cross-section $\sigma_{160} = 5.1 \times 10^{-26}$cm$^2$ per H-atom above, 
the total H column density at each position is
\begin{equation}
{\rm N_{Htot}} = S_{160 \mu m}  \frac{1}{\sigma_{160_{\mu m}}\ B_{{160\mu m},{\rm T_d}}}
\label{eq.160mu}
\end{equation}
so that the H$_2$ column density is simply 
\begin{equation}
{\rm N_{H_{2}}} = \frac{1}{2}  ({\rm N_{Htot}} - {\rm N_{\ion{H}{I}}})
\end{equation}
The left panel of Fig. \ref{fig.NH2_160mu} shows the total H$_2$ column density derived in this way {  using an ${\rm N_{\ion{H}{I}}}$ map smoothed to the 160\mum 40$\arcsec$ resolution.}

In Table 2 the H$_2$ mass estimates for individual clouds are only based on the 8${\rm \mu m}$  map due to angular resolution -- the 160 \mum data is at a resolution larger than the clouds.  Tables 3 and 4, which provide a summary of the molecular gas mass calculations, include the 160 \mum results.

It is very difficult to estimate uncertainties for our mass estimates.  Statistical noise related errors, as manifested by the variations within ``blank" areas of maps, are about $10^{20}$ in column density.  Systematic uncertainties are certainly present as well so we consider the column density noise in our maps to be $\sim2 \times 10^{20}$ H$_2$ cm$^{-2}$.

%--- Compute cross-section\\ 

% $$\sigma_{160_{\mu m}}=S_{160_{\mu m}} \frac{1}{B_{160\mu m}(T_d)\   \NHI}$$

%--- Use formula\\
%$$2\  \NHmol= \left(\frac{\NHI}{\sigma_{8\mu m}}\right)_{o}\sigma_{8\mu m} - \NHI$$
\subsection{The $\ratiot$ ratio in NGC 6822}
We have estimates of the CO-to-${\rm H_2}$ conversion factor on two different scales, at the cloud level and for the whole area mapped by HERA in CO(2--1).
%\begin{table}[tbp]
%\caption{Average over the cloud sample of the $\ratiot$ factor for the different methods used to derive the molecular gas mass. Averages for the by eye and {\tt CPROPS} methods are respectively over 14 and 9 clouds. In the case of the virial masses the number of clouds in the sample are 17 and 11 respectively for the ``by eye'' and {\tt CPROPS} methods (see Table \ref{tab.clouds}).}
%\label{tab.Xfactor}
%\begin{flushleft}
%{\renewcommand{\arraystretch}{1.2}
%\begin{tabular*}{88mm}{@{\extracolsep{\fill}}llrr} \hline \hline
%\multirow{2}{*}{Method}& &\multicolumn{1}{r}{ $\ratiot$} & Range \\ 
%& &\multicolumn{1}{r}{ ${\rm cm^{-2}/K\kms}$} &${\rm cm^{-2}/K\kms}$\\ \hline
%\multirow{2}{*}{${\rm 8\mu m}$} &  by eye & $1.9\pm0.8\times10^{21}$ & $0.8-3.0\times10^{21}$\\
%& {\tt CPROPS} & $1.9\pm1.0\times10^{21}$& $0.7-3.8\times10^{21}$ \\ \hline
%${\rm 160\mu m \mbox{ with }}$ & by eye & $1.7\pm0.9\times10^{21}$ & $0.4-3.2\times10^{21}$ \\
%${\rm T_{dust}\mbox{ map }}$& {\tt CPROPS} & $1.4\pm1.0\times10^{21}$ & $0.3-3.0\times10^{21}$\\ \hline
%%${\rm 160\mu m \mbox{ with }}$  & by eye & $1.9\pm1.2\times10^{21}$ & $0.04^{\mathrm{a}}-4.19\times10^{21}$\\
%%${\rm T_{dust}=19K}$& {\tt CPROPS} & $2.1\pm1.4\times10^{21}$ & $0.33-5.18\times10^{21}$\\ \hline
%\multirow{2}{*}{${\rm M_{vir}}^{\mathrm{}}$} & by eye & $2.2\pm3.2\times10^{21}$ & $0.2-12\times10^{21}$\\
%& {\tt CPROPS} & $1.4\pm1.0\times10^{21}$ & $0.3-3.1\times10^{21}$\\ \hline
%\end{tabular*}}
%\end{flushleft}
%\end{table}

\begin{table}[tbp]
\caption{Average over the cloud sample of the $\ratiot$ factor for the different methods used to derive the molecular gas mass.}
\label{tab.Xfactor}
\begin{flushleft}
{\renewcommand{\arraystretch}{1.2}
\begin{tabular*}{88mm}{@{\extracolsep{\fill}}llrr} \hline \hline
\multirow{2}{*}{Method}& &\multicolumn{1}{r}{ $\left<\ratiot\right>$} & Range \\ 
& &\multicolumn{1}{r}{ ${\rm cm^{-2}/K\kms}$} &${\rm cm^{-2}/K\kms}$\\ \hline
\multirow{2}{*}{${\rm 8\mu m}$} &  by eye & $2.1\pm0.8\times10^{21}$ & $0.3-4.3\times10^{21}$\\
& {\tt CPROPS} & $3.5\pm1.4\times10^{21}$& $1.0-6.4\times10^{21}$ \\ \hline
${\rm 160\mu m \mbox{ with }}$ & by eye & $1.5\pm0.9\times10^{21}$ & $0.5-3.1\times10^{21}$ \\
${\rm T_{dust}\mbox{ map }}$& {\tt CPROPS} & $1.3\pm0.9\times10^{21}$ & $0.4-2.7\times10^{21}$\\ \hline
%${\rm 160\mu m \mbox{ with }}$  & by eye & $1.9\pm1.2\times10^{21}$ & $0.04^{\mathrm{a}}-4.19\times10^{21}$\\
%${\rm T_{dust}=19K}$& {\tt CPROPS} & $2.1\pm1.4\times10^{21}$ & $0.33-5.18\times10^{21}$\\ \hline
\multirow{2}{*}{${\rm M_{vir}}^{\mathrm{}}$} & by eye & $1.6\pm0.8\times10^{21}$ & $0.2-2.8\times10^{21}$\\
& {\tt CPROPS} & $1.1\pm0.8\times10^{21}$ & $0.07-3.5\times10^{21}$\\ \hline
\end{tabular*}}
Averages for the ``by eye'' and {\tt CPROPS} methods are respectively over 11 and 9 clouds. In the case of the virial masses the number of clouds in the sample are 12 and 11 respectively for the ``by eye'' and {\tt CPROPS} methods (see Table \ref{tab.clouds}).
\end{flushleft}
\end{table}

\begin{table}[tbp]
\caption{Molecular gas masses, $\ratioo$ conversion factor and characteristic time to transform molecular gas into stars derived from IR emission.}
\label{tab.masses}
\begin{flushleft}
\begin{tabular*}{88mm}{@{\extracolsep{\fill}}lrrrr}
\hline
\hline
	& total ${\rm M_{H_{2}}}$ &map ${\rm M_{H_{2}}}$ & $\ratioo$ & $\tau$\\
	& ${\rm M_{\sun}}$& ${\rm M_{\sun}}$ & ${\rm \frac{cm^{-2}}{{K\kms}}}$ & yr \\[1ex]
\hline
\noalign{\smallskip}
${\rm 8\mu m}$ & $9.2\times10^6$ &$5.7\times10^6$ & $5.3\times10^{21}$ & $2.2\times10^8$\\
%$160{\rm \mu m}$ &$<1.6\times10^{7}$ &$<1.0\times10^{7}$& $<9.2\times10^{21}$ & $<3.8\times10^8$  \\
%(19K) & & & &\\
%$160{\rm \mu m}$ &${<7.3\times10^{6}}$ &$5.5\times10^{6}$& $5.1\times10^{21}$ & $2.1\times10^8$  \\
%(20K) & & & &\\
$160{\rm \mu m}$& ${4.6\times10^{6}}$ & $3.4\times10^{6}$&$2.3\times10^{21}$ & $1.3\times10^8$ \\
\hline
\end{tabular*}
The second column refers to the area mapped in CO. The last column is the inverse of the SFE.
\end{flushleft}
\end{table}

Table \ref{tab.Xfactor} presents averages and total ranges of the $\ratiot$ factor for our sample of clouds for the different methods we have used. The average is over 9 clouds in the case of {\tt CPROPS} and 14 in the case of the ``by eye'' identification. For the virial mass the average is over our full sample of 11 and 15 clouds using a projected area as defined in Sec. \ref{sec.CPROPS}. We have included the virial mass as a valid method to estimate cloud masses because the size-linewidth relationship we find (see Fig. \ref{fig.size_width}) is similar to the one for Galactic clouds. This suggests that the CO molecules are not greatly photodissociated at the outer edge of the clouds and that the cloud size as determined from the CO emission is close to the true size of the molecular clouds.  As might be expected in this case, the masses computed by the virial theorem are similar to but usually lower than  the ones we estimate by the other methods (see Table \ref{tab.clouds}).

We find similar values of $\ratiot$ for the different methods used, around 1.5--$2\times10^{21}\Xunit$, 5--10 times the $\approx 2\times10^{20}\,\Xunit$ value for the inner part of the Milky Way in CO(1--0) \citep[e.g. ][]{Dickman:1986}. From interferometric CO data \citet{Bolatto:2008} using only virial masses for a sample of dwarf galaxies of the Local Group find a value of the CO-to-${\rm H_2}$ factor similar to the Galactic value despite the low metallicities.   Presumably this is because they detect the dense protected parts of bright clouds.
%However, \citet{Blitz:2006} with similar methods and data find higher $\ratioo$ values for low metallicity systems.

There has been evidence for a long time that as the size scale increases, at least for low metallicity objects, the $\ratioo$ factor increases as well \citep{Rubio:1993}.
Table \ref{tab.masses} shows the mass estimates for the area of NGC~6822 mapped by HERA and a larger region including the HERA map and which covers virtually all of the stellar and H$\alpha$ emission.  The $\ratiot$ factors and characteristic times to transform molecular gas into stars are shown in columns 3 and 4 for the different methods used to estimate the molecular gas column density. 
%The total molecular gas masses over the whole galaxy (col. 1) are upper limits because 
After subtraction of the HI column density, some pixels became negative; we have attributed a nil value to the pixels in the N${\rm H_2}$ map where the molecular gas column density had a negative value after application of Eq. 4. The $\ratiot$ factor for the whole of the area mapped in CO(2--1) is slightly larger ($\sim5\times10^{21}\,\Xunit$) than for the individual clouds and 20--25 times larger than the standard Galactic value for the molecular ring of the Milky Way. This higher value is close to the 4 -- 8 $\times10^{21}\,\Xunit$ found by \citet{Israel:1997} for scales of $\sim50\arcsec$.
We now turn to modeling to better understand the relationship between CO emission, H$_2$ mass, and the other properties of the clouds.

%
%Standard Galactic factor : $2\times10^{20}cm^{-2}$ \\
%From 8mu CPROPS $\langle X \rangle=3.53\times10^{21}\ {\rm cm^2}$ $
%\sigma=2.13\times10^{21}\ {\rm cm^2}$ \\
%From 8mu A la main  $\langle X \rangle=3.12E+21\times10^{21}\ {\rm cm^2}$ $\sigma=2.38\times10^{21}\ {\rm cm^2}$ \\
%Several estimates:\\
% 8 mu, Virial eye, Virial {\tt CPROPS}, 160mu with temp hypothesis.\\
%Is there H$_2$ without CO? \\
%In M31 CO correlates with extinction.\\
%In following section, lowest kin temp corresponds roughly to lowest $\ratioo$ value.\\

\section {Modeling the CO emission from NGC~6822 GMCs}

\begin{figure}[tp]
\begin{flushleft}
\includegraphics[angle=0,width=88mm]{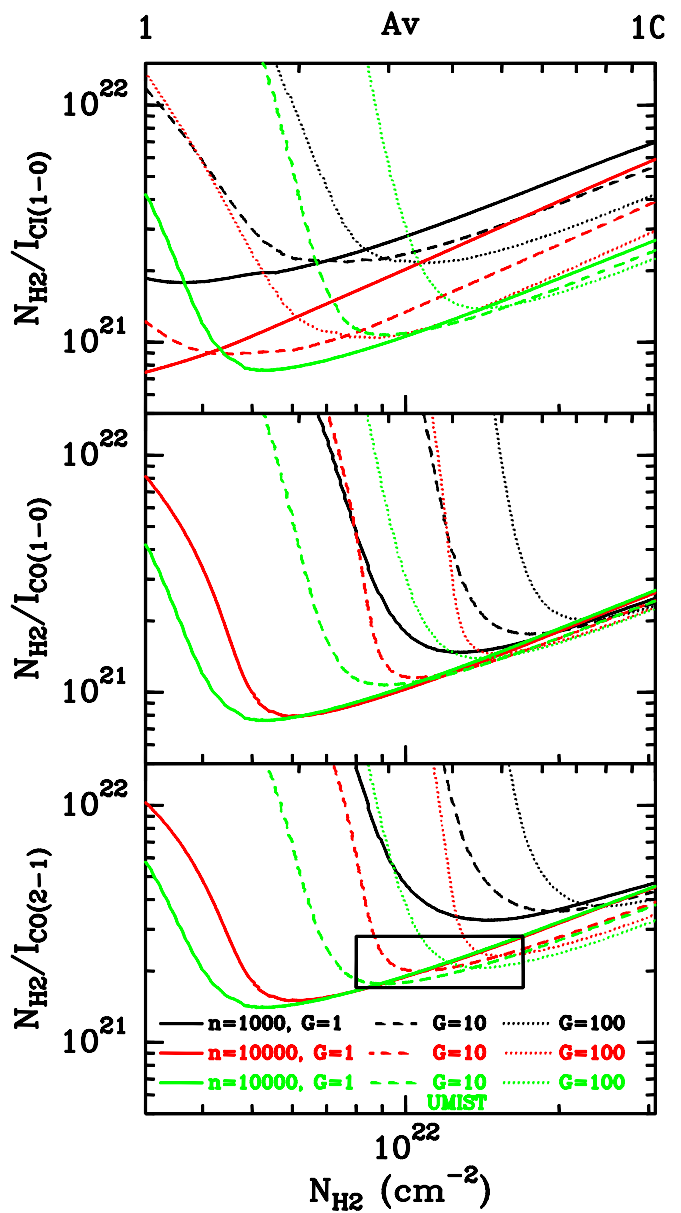}
\caption{Model $N{\rm(H_2)}/I_{\rm \ion{C}{I}}$ (top) and 
$\ratioo$ factors as a function of visual extinction A$_v$
and H column density into the cloud.
As the atomic Hydrogen skin is very thin even for the
highest UV field used here, the clouds are essentially
completely molecular so the depth is shown as H$_2$ depth
not H.
The $x-$axis extends beyond standard cloud column densities
in order to show how a change in cloud structure  with
respect to Galactic clouds would affect the CO intensities.
The expected range is roughly indicated by a box in the
lower panel.
While the UV field has little effect on the Hydrogen, it has
a major effect on the C$^+$/C/CO transition.}
\label{fig.NH2}
\end{flushleft}
\end{figure}

\begin{figure}[tp]
\begin{flushleft}
\includegraphics[angle=0,width=88mm]{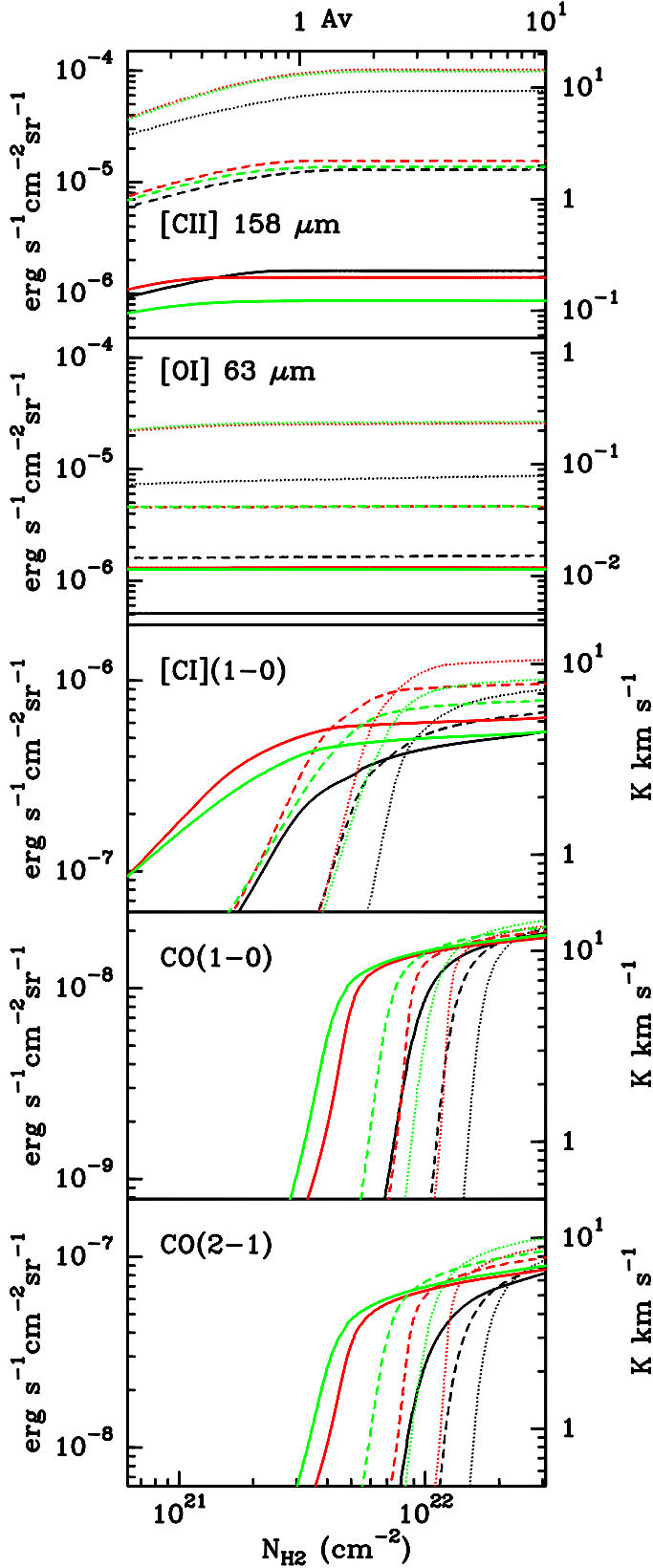}
\caption{Line strengths of the major observables \ion{C}{II}, \ion{O}{I}, \ion{C}{I}, and CO
as a function of visual extinction A$_v$ and H column
density into the cloud.   Lines are as in Fig. 8 -- black
for density $n=1000$, red for the nominal $n=10000$ with
CLOUDY default rates, and green for $n=10000$ with UMIST
rates.  The solid, dashed, and dotted lines are for
respectively UV fields of 1, 10, and 100 times the estimated
solar neighborhood value.}
\label{fig.energy}
\end{flushleft}
\end{figure}

\subsection{Description  of the models}
We have modeled the structure and the emission of the molecular
clouds in NGC~6822 using CLOUDY version 07.02 \citep{Ferland:1998}.
CLOUDY  computes the spectrum of a gaseous nebula using as only
inputs the geometry, the gas and dust composition and the energy
input on the nebula. Originally designed to deal with photoionized
nebula, CLOUDY is now able to perform accurate calculations in the
Photo Dissociation Region (PDR) and even well into the molecular
clouds due to the implementation of the H$_2$ physics
\citep{Shaw:2005} and a network of 1000 reactions involving 68
molecules \citep{Abel:2005}. CLOUDY determines steady-state solutions
for the chemistry and predicts the emission in the fine structure
lines and CO, which are the main coolants in the PDR and the
molecular cloud.
%JB H$_2$ ???

The exact geometry of the regions to be modeled, the position of
the excitation sources and the shape of the incident continuum
are not well known. Therefore we have performed plane-parallel
models with the incident continuum determined by \cite{Black:1987}
for the local interstellar radiation field (ISRF). This continuum
is not restricted to the FUV range as in many classical PDR
calculations. To estimate the intensity of the incident continuum
we have taken into account that according to \citet{Draine:2007a}, the
gas in NGC~6822 is exposed to a minimum radiation field of order
twice the local ISRF. On the other hand, the variation in the
GALEX FUV observations of NGC~6822 is about a factor 50 from the
general field to the stronger fields around specific positions.
Therefore, we have computed simulations for scaled versions of the
local ISRF by factors of 1, 10, and 100. However, most of the
clouds in our regions of interest are exposed to (GALEX-based)
fields of less than 10.  In addition to the \cite{Black:1987}
continuum, the calculations include the cosmic microwave
background and a cosmic-ray ionization rate of $2.5 \ 
10^{-17}$s$^{-1}$, which is typical of the Milky Way
\citep{Williams:1998} but not known for NGC~6822.
%JB
Given that UV photons penetrate further into molecular clouds in the low-metallicity environment of NGC 6822 than into their Galactic counterparts, the cosmic ray flux
is not as critical a parameter.

We have adopted typical gas abundances and dust distributions
(including PAHs) of the Milky Way interstellar medium (see CLOUDY
documentation for the exact values). These abundances are scaled
using the metallicity of NGC 6822 (0.3 solar). The
simulations have been computed for two densities, 1000 and 10000
%JB mol
mol cm$^{-3}$, which are representative of the bulk of the molecular
gas. In addition we have computed simulations with two series of
rate coefficients -- the CLOUDY default and the UMIST. The reason
for testing two sets of rate coefficients is that in the
comparison of PDR codes \citep{Rollig:2007}, the \ion{C}{I}--CO transition
was found to vary significantly (from $A_v \sim 2$ to $A_v \sim
5$) depending largely on these coefficients.  This is a critical
region for low-metallicity molecular clouds because the CO
emission depends strongly on the fraction of the cloud in which CO
has formed.

In summary, we have explored a grid of 12 models (3 fields
$\times$ 2 densities $\times$ 2 reaction rates). In all the
simulations, a line width of 2 km/s is included via the {\tt
turbulence} command. The results are summarized in 
Figs. \ref{fig.NH2} and \ref{fig.energy}. In
both figures, the simulation results are shown for three
radiation fields, solid lines for $G_0=1$, dashed for $G_0=10$ and
dotted for $G_0=100$ (where $G_0$ is the intensity of the far
ultraviolet continuum in the cloud surface in units of the
\cite{Habing:1968} field). The red and black curves use the CLOUDY
default rate coefficients for respectively densities of $10^4$ and
$10^3$ cm$^{-3}$.  The green curve uses the UMIST rate
coefficients for $10^4$ cm$^{-3}$.

The range of H$_2$ column densities explored goes from $\sim 3 \  10^{21}$~cm$^{-2} $ to $\sim 3 \ 10^{22}$~cm$^{-2}$.
Assuming that the grain abundance scales with metallicity, those column densities
correspond to visual extinctions ($A_v$) from 1 to 10.
Therefore, the model simulations cover the full range of H$_2$ 
column densities expected in the region of interest, from a few $10^{21}$~cm$^{-2}$  of the extended component derived from the dust emission to $\sim 10^{22}$~cm$^{-2}$
towards the center of the molecular clouds detected in CO (see Fig. \ref{fig.NH2_8mu}).
The latter value is also typical for molecular clouds in the Milky Way 
\citep{Larson:1981}, although regions of higher column density also exist, 
particularly towards cloud cores and GMC centers.  

Among the goals of the modelling is to see if the $\ratioo$ factor
is close to our other estimates  and to predict testable fluxes in
other important cooling lines which could then act as further
diagnostics. 
Figure \ref{fig.NH2} shows the $\ratioo$ and $\ratioc$ 
ratios as a function of the total column density and visual
extinction into the cloud for all models. In other words, at a
given depth into the cloud the plotted $\ratioo$ and $\ratioc$
ratios are the integrated values from the cloud surface to that cloud depth.
 While, Fig. \ref{fig.energy} shows the cumulated intensity for 
different lines as as a function of the total column density  into the cloud. 

The column densities shown in Figs. \ref{fig.NH2} and \ref{fig.energy} are
actually total (H + H$_2$) but the gas is almost completely
molecular due to the low radiation fields (the model with a
highest atomic H fraction has only 3 $\%$ of H in the region of
interest).

The \ion{C}{I} to CO transition occurs when the lines in Fig.  \ref{fig.NH2} are 
decreasing steeply due to the fact that CO begins to emit strongly just after
formation. This is clearer in the two lower panels of Fig. \ref{fig.energy},
which shows how the energy emitted is divided into the principal
PDR cooling lines: CO, \ion{C}{I}, \ion{C}{II}, and \ion{O}{I}[63]. 
The models suggests that no CO emission is expected for H$_2$
column densities below $\sim 2$\,10$^{21}$ cm$^{-2}$.

\subsection{Constraining the physical conditions and the $\ratioo$ ratio}

Currently, the spectral data available to study NGC 6822
is limited to HI and CO. However, in this section we will try to
constrain the physical conditions and the $\ratioo$ ratio using our model
calculations and assuming line ratios typical of normal galaxies. 
In doing so, we will also make use of GALEX measurement of the far-UV field
in NGC 6822 \citep{Draine:2007a}. We will also take into account
our previous estimations of the H$_2$ column density of the 
CO clouds detected in NGC 6822. The exact value is not important since we show 
below that the  $\ratioo$ conversion factor depends weakly on NH2 in the region of interest. Therefore, the $\ratioo$ conversion factor derived in this section
remains independent of other previous estimations.

For normal spiral galaxies, the \ion{C}{II}[158] integrated intensities in
energy units are typically 1000 -- 2000 times the CO(1-0)
intensity \citep {Stacey:1991,Braine:1999}. On the other hand, the \ion{O}{I}[63]
to \ion{C}{II}[158] ratio is $\sim 0.3$ in normal galaxies at large scales
\citep{Braine:1999, Lord:1996,Malhotra:2001} and can be even higher when
studying smaller scales as the spiral arms of M31
\citep{Rodriguez-Fernandez:2006}. These are lower limits to the \ion{O}{I}/\ion{C}{II} ratios
in the neutral components of the ISM since a fraction of the
observed \ion{C}{II} emission may come from ionized gas not closely
associated with molecular clouds or PDRs.

The simulations show that the \ion{C}{II} flux varies linearly
with the incident field and depends little on the density for the
range explored here, as expected. The \ion{O}{I} flux depends strongly on
both density and radiation field and for the low density models
the flux is well below the large-scale value of 1/3 that of the
\ion{C}{II} typical of normal galaxies. Thus, a density of 10$^4$ cm$^{-3}$ is
more appropriate to reproduce the \ion{O}{I}/\ion{C}{II} ratio.

Even for large column densities where the CO is formed for all the
parameters studied, the simulations predict strong variations in
the {\ion{C}{II}/CO(1-0)} ratio from 100 to $10^4$ for incident field of
$G_0=1$ to 100 respectively. To reproduce ratios of 1000-2000
typical of normal spiral galaxies a relatively low intensity
incident field of $G_0=10$ is favored. Similarly, the model
{\ion{C}{I}/CO(1--0)} ratio also favors low intensity fields since the
predicted ratios are high, about 0.5 in K km/s except for $G=1 $
with a ratio of about 0.3, whereas typical observed values are
about 0.2 \citep[e.g.][]{Gerin:2000} and often less.  For lower
columns or the late CO formation curves, the \ion{C}{I}/CO ratio is
substantially higher (due to the weak CO).

All together, the comparison of the model prediction with the
typical line ratios yields a representative density of $\sim$ 10$^4$ cm$^{-3}$ and
incident field in the range 1--10, in agreement with
\cite{Draine:2007a} and our GALEX estimates for NGC~6822.  
Taking into account those density and incident radiation fields, 
and a range of H$_2$ column densities from 8\,10$^{21}$ cm$^{-2}$
to 2\,10$^{22}$ cm$^{-2}$, which comprise the expected H$_2$ column density
of the CO clouds detected in NGC 6822, 
a likely value for $\ratiot$ is in the range
$\sim 2-3 \times 10^ {21}\Xunit$. 
These estimations of the $\ratiot$ factor do not depend
critically on the exact H$_2$ column density of the molecular clouds in NGC 6822 and agree reasonably well with other estimations. 
Increasing the radiation field further pushes the CO edge of the cloud deeper in,
probably unrealistically deep unless the molecular clouds in NGC
6822 have higher column densities than Galactic GMCs. Observing
the \ion{C}{II}, \ion{O}{I}, and \ion{C}{I} lines with the Herschel satellite would be of great interest because they are sensitive respectively to the SFR,
the SFR and density, and the cloud depth (the others being
independent of total cloud column density -- cf. Fig.  \ref{fig.energy}).

\section{The efficiency of star formation in NGC~6822}

Using the ${\rm H\alpha}$ luminosity  and the calibration from \citet{Kennicutt:1998}, \citet{Cannon:2006} derive a global star formation rate of $0.015\, {\rm M_{\sun}~yr^{-1}}$.  \citet{Israel:1996} find $ 0.04\,{\rm M_{\sun}~yr^{-1}}$ over the last $10^7$ years
from bolometric luminosity measurements. Using the calibration from \citet{Hunter:1986} which supposes that close to half of the ionizing photons are lost to dust and a Salpeter IMF:
\begin{equation}
{\rm \dot{M}}=5\times10^{-8} L(H_{\alpha})/L_{\sun}~{\rm M_{\sun}~yr^{-1}}
\end{equation}
yields, with ${\rm L(H_{\alpha})}=2\times10^{39}~{\rm erg~s^{-1}}$ \citep{de-Blok:2006a} a star formation rate of $ 0.026 \,{\rm M_{\sun}~yr^{-1}}$; this is the value we will use for NGC~6822.\\
%QUESTION: is this the Halpha lum of Cannon?\\
%ANSWER: Yes Cannon uses deBlok's value \\
The star formation efficiency is usually defined as the ratio of the star formation rate over the mass of molecular gas available to form stars:
\begin{equation}
{\rm SFE=\frac{SFR}{M_{H_{2}}}}
\end{equation}
The characteristic time to transform molecular gas into star is directly $\tau={\rm 1/SFE}$. Table \ref{tab.masses} shows the values of the characteristic time $\tau$ for the different methods used to estimate the molecular gas column density. The H$_2$ consumption times are smaller than those found for large spirals, about ${\rm 2\times10^9~yr}$ \citep{Murgia:2002,Kennicutt:1998}, such that the SFE in NGC 6822 is higher by the same ratio. This is consistent with the high star formation efficiency found by \citet{Gardan:2007} in M~33, a spiral galaxy larger than NGC~6822 but an order of magnitude less luminous and less massive than the Milky Way.  Two other small galaxies, IC10 and NGC~2403 (which is very similar to M33), have also been found to have high SFEs by respectively \citet{Leroy:2006}, \citet{Kennicutt:1998} and \citet{Blitz:2007} but the constraints on the $\ratioo$ conversion are not 
stringent.

Going to $z\sim 1$, the star formation rate increases by at least one order of magnitude \citep{Madau:1996,Wilkins:2008}. Even if the gas fraction was higher in the past, a high star formation efficiency has to be introduced in order to explain such a wide variation of the star formation rates. Small Local Group galaxies such as NGC~6822, M~33, the LMC or the SMC share some properties with intermediate redshift objects: they are gas rich, have subsolar metallicities and seem to exhibit high star formation efficiencies. {  Low-luminosity NGC 6822 also shares with early universe objects  a high FIR/CO luminosity ratio.
Note that both CO and the dust emission are affected by metallicity.
However, unlike these rare but very luminous galaxies, NGC~6822 (like the other small local objects above)
has a low SFR, placing it in an empty region of Figures 8 and 9 in \citet{Solomon:2005}}. High spatial resolution observations of these {  local} systems may help us understand the physics of intermediate redshift galaxies.
%  First address the SFR using the previously used tracers: Halpha, FUV, FIR, 8mu ...
%  Cannon -- 0.015 Moyr from dBW06 Halpha + Kenn98  calibration.
%  try calzetti 24 + Ha cal
%  but 0.04 by Gallart 96 over last 100-200 Myr.
% Add Gallart and possibly Calzetti values and refs

\section{Hubble X and Hubble V}
\subsection{Diffuse CO emission South of Hubble X}
\label{sec.diffuse}
\begin{figure}[tp]
\begin{flushleft}
\includegraphics[angle=270,width=88mm]{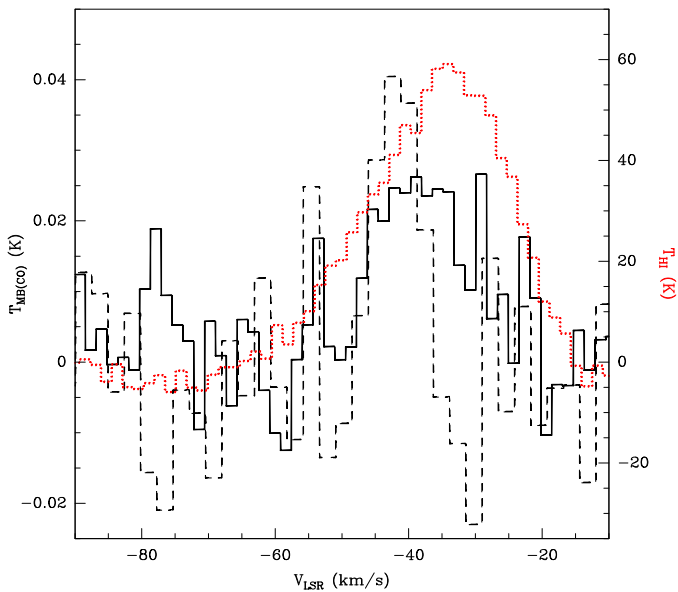}
\caption{CO(2-1) and \ion{H}{I} spectra of the region south of of Hubble X in
which CO emission appears to be detected (cf. Fig. \ref{fig.COmom0}).
The solid line is the CO(2--1) summed over a 2.57 arcmin$^2$ region east of Hubble X, the dotted
line is the \ion{H}{I} 
spectrum summed over the same region, and the 
dashed line is a JCMT CO (2--1) spectrum taken at
 RA~$19^\mathrm{h}42^\mathrm{m}19\ \fs7$, Dec~$-14\degr50\arcmin31\arcsec$.}
\label{fig.diffuse}
\end{flushleft}
\end{figure}
%\begin{figure}[tp]
%\begin{flushleft}
%\includegraphics[angle=270,width=88mm]{HubX.eps}
%\caption{CO(2--1) and \ion{H}{I} emission from Hub X CO data from Israel -- DETAILS !!}
%\label{fig.HubX}
%\end{flushleft}
%\end{figure}

\begin{table}
\caption{Properties of the \ion{H}{II} regions Hubble V and Hubble X.}
\label{tab.HubVandX}
\begin{tabular*}{88mm}{@{\extracolsep{\fill}}lrrrr}
\hline
\hline
Name & $\left.\alpha_{\mathrm{off}}\right.^{\mathrm{a}}$ & $\left.\delta_{\mathrm{off}}\right.^{\mathrm{a}}$ & ${\rm T_{CO\ mb}}$ & ${\rm I_{CO}}$ \\
     & $\arcsec$ & $\arcsec$ & mK & ${\rm K \kms}$ \\
\hline
Hubble V &$-68.5 $&$299.0$ &500 &$0.93$  \\
Hubble X &$55.2$&310.8 &$\left.<85\right.^{\mathrm{b}}$& $\left.<0.19\right.^{\mathrm{c}}$ \\
\hline
\end{tabular*}
\begin{flushleft}
\begin{list}{}{}
\item[$^{\mathrm{a}}$] Offsets with respect to the reference position\linebreak $(\alpha_{o},\delta_{o})=(19^\mathrm{h}44^\mathrm{m}57\ \fs8,-14\degr48\arcmin11\arcsec )$
\item[$^{\mathrm{b}}$] Corresponding to a $3\sigma$ level 
\item[$^{\mathrm{c}}$] Corresponding to a $3\sigma$ level and a $3\kms$ fwhm linewidth.
\end{list}
\end{flushleft}
\end{table}

A major question is whether large quantities of H$_2$ could be missed through the use of CO as a tracer of molecular gas.  In particular, does the presence of the luminous star forming region Hubble X (see Table \ref{tab.HubVandX} for positions of Hubble V and X) with copious H$\alpha$, FUV, 160$\mu$m and 8$\mu$m emission but without a CO detection invalidate CO as a tracer of H$_2$ in galaxies like NGC~6822?  As shown in Fig. \ref{fig.NH2_8mu}, for reasonable estimates of the radiation field or dust temperature, the dust emission corresponds to that expected from the atomic component alone.  Thus, we believe that towards the \ion{H}{II} region there is in fact very little H$_2$, and not just little CO.  However, we see evidence for ``diffuse'' CO emission from the \ion{H}{I}-rich region near and South of Hubble X (see Fig. \ref{fig.diffuse}).  

In Fig. \ref{fig.diffuse}, we have summed the CO and \ion{H}{I} spectra over a region roughly $50\arcsec \times 200\arcsec$ in size, yielding an apparent detection in CO despite the absence of individual clouds with detectable CO emission.  Also shown is the result of an observation towards $19^\mathrm{h}42^\mathrm{m}19\ \fs7$, $-14\degr50\arcmin31\arcsec $ made with the JCMT in CO(2--1) at $21\arcsec$ resolution in 1992, which, although the signal-to-noise ratio is low, seems to show CO emission towards roughly the same position and at the same velocities.  With the resolution available in 
extragalactic observations, ``diffuse'' signifies that neither spatially nor spectrally 
can we distinguish the clouds which, taken together over a large area, appear 
to contribute a detectable CO signal.  While these clouds could be like Galactic cirrus,
our resolution and brightness sensitivity are not sufficient to be sure.  They are {\it not} like either galactic GMCs or the individual clouds discussed in Sections \ref{sec.clouds}.  Using the methods described in Sect. 6, no H$_2$ is found in this area although the CO luminosity is $\sim 10^4 $~K~km/s~pc$^2$.

\subsection{The $^{13}$CO and HCN emission in Hubble V} 
\label{sec.HubV}
\begin{figure}[tp]
\begin{flushleft}
\includegraphics[angle=0,width=88mm]{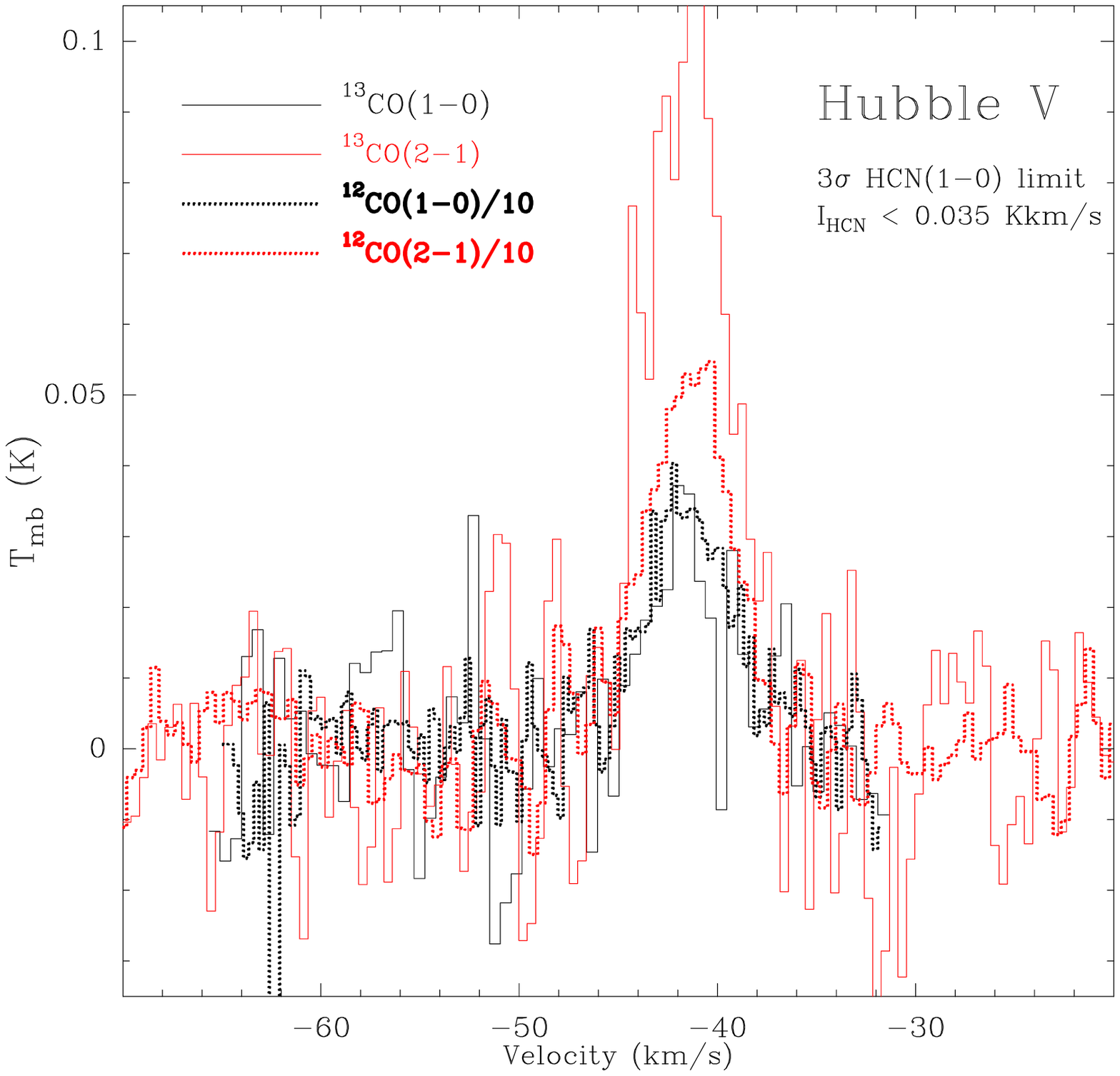}
\caption{Spectra of the Hubble V \citep{Hubble:1925} \ion{H}{II} region.  (\emph{dotted line})~~\inst{12}CO divided by 10, (\emph{continous line})~~\inst{13}CO, (\emph{black}) $\mathrm{J}=2\rightarrow1$, (\emph{red}) $\mathrm{J}=1\rightarrow0$.}
\label{fig.HubV}
\end{flushleft}
\end{figure}

Unlike most spiral galaxies 
%SAKAMOTO FUKUI 2003-2005 
\citep{Braine:1993}, comparing the CO(1--0) emission with the CO(2--1) emission convolved to $21\arcsec$ angular resolution
yields a line ratio of about ${\rm {CO(2-1)}/{CO(1-0)} \sim 1.2}$ in Hub V.  This is unlikely to be the case over much of NGC~6822 because rather warm and dense gas is required, including some optically thin emission.  
The \element[][13]{C}O emission confirms this for Hubble V -- the \element[][12]{C}O/\element[][13]{C}O ratio is 15 in the (1--0) transition
but only 5.7 in the (2--1) transition, showing that the higher transition is efficiently excited.  
\citet{Israel:2003} found considerably higher \element[][12]{C}O/\element[][13]{C}O line ratios with the SEST telescope. This is a strong indication for the presence of tenuous molecular gas
with low CO optiocal depths surrounding the Hubble V cloud, probably
similar to the diffuse gas found near Hubble X (see Sect. \ref{sec.diffuse})
% See also (Israel ...for other lines ..).
%Situation in LMC and SMC at large scales ?

In order to estimate the fraction of dense gas in Hubble V, we observed the HCN(1--0) line at 88.6316 GHz.
HCN has a high dipole moment and thus requires high densities to be excited.
HCN(1--0) was not detected despite reaching a $3 \sigma$ noise level of 0.035 K km s$^{-1}$, a factor 60 below the CO(1-0) line.  Typical values of CO(1--0)/HCN(1--0) are $\sim 50$ in galactic disks (\citealp[for M~51]{Kuno:1997}, \citealp[for NGC 4414]{Braine:1997}, \citealp[for M~31]{Brouillet:2005}),
10 -- 20 in nuclei \citep{Nguyen:1992, Henkel:1991}, and less in ultraluminous infrared galaxies which are as a result believed to
have a particularly high fraction of their H$_2$ in dense cores \citep{Gao:2007}.
Again with the SEST, \citet{Israel:2003} reports detection of the {{HCO$^+$(1--0)}} line in Hubble V, at 4 times the brightness of our limit to {HCN(1--0)}.  While rather extreme, \citet{Brouillet:2005} noted an apparent rise in the HCO$^+$/HCN
ratio going towards the outskirts of M31, where the metallicity presumably decreases.
From the data presented in this paper, despite being a major \ion{H}{II} region, Hubble V is not particularly rich in dense gas.
If the CO underestimates the H$_2$ mass in Hubble V, then the dense gas fraction in Hubble V is
lower than in spiral disks 
because the HCN in dense cores should be less affected by the dissociating radiation field than the CO. 

\section{Galactic emission}
\label{sec.cirrus}

\begin{figure}[tbp]
\begin{flushleft}
\includegraphics[angle=270,width=88mm]{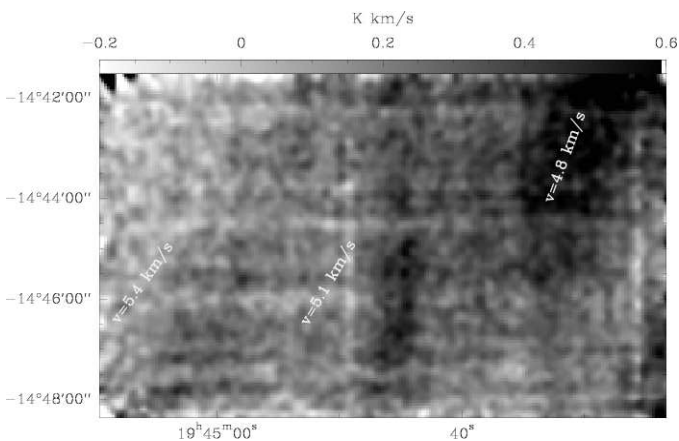}
\caption{CO(2--1) emission from the local molecular ``cirrus'' cloud along the line of sight to NGC~6822.  Indicated on the figure are the velocities of the emission at 3 different positions.  }
\label{fig.cirrus}
\end{flushleft}
\end{figure}

A factor which greatly complicates studies of NGC~6822 is the presence of a Galactic molecular cloud along the line-of-sight towards NGC~6822.  Some optical/UV emission is absorbed and the cloud emits at (at least) FIR wavelengths, making it difficult to clearly identify the emission coming from NGC~6822.  This is particularly a {  question} at 160$\mu$m where the dust temperature apparently decreases to the West but this is likely due to the increasing column density and low temperature of the Galactic cloud.  In CO, the Galactic cloud can be separated from NGC~6822 (see Fig. \ref{fig.cirrus}) so that we know the western part is more affected.  A velocity gradient from roughly East to West is present in the cloud, in addition to the column density gradient.  For a cloud distance of 100 pc, realistic due to the $-20\degr$ galactic latitude of NGC~6822, the resolution is about 1500 a.u. or 0,007 pc, making this one of the highest-resolution observations up to now of a local cloud.
A dedicated study of the large scale structure of this Galactic cirrus cloud will be presented by Israel et al. (in prep). {  Figure \ref{fig.cirrus} shows where FIR emission and optical/UV absorption are most expected.  The decrease in dust temperature at the NW corner of the 160$\mu$m map in Fig. \ref{fig.NH2_160mu} might be due to the cirrus.  However, from looking at the region around RA 19:44:32 Dec -14:45:00, it is clear that any emission from the cirrus falls below our intensity cuts, whether at 100 or 160$\mu$m.} 

\section{Conclusions}
From large-scale CO mapping of a very nearby low metallicity galaxy, we identify a sample of molecular clouds, most of which were also detected by the CPROPS algorithm, allowing an unbiased assessment of their properties.  The properties of these GMCs (size, linewidth, virial mass) appear similar to Galactic clouds but the H$_2$ mass per CO luminosity is much higher such that we estimate $\ratiot \approx 2 \times 10^{21}\Xunit$ for the clouds.  A variety of methods yield coherent values.
Our modeling with the CLOUDY program also provides similar ratios for reasonable parts of parameter space.  At large scales ($\sim$ kpc, well above that of the clouds), we find evidence for a higher $\ratioo$, $\sim 4 \times 10^{21}\Xunit$ , in agreement with work in NGC~6822 by \citet{Israel:1997} and in the SMC by \citet{Rubio:1993}.  No evidence for H$_2$ in the absence of CO emission was found. The molecular gas masses derived using the high $\ratioo$ values estimated here, coupled with an ${\rm SFR = 0.026\,M_{\sun}\,yr^{-1}}\pm50\%$, support the idea that molecular gas is more quickly cycled into stars in these small low metallicity galaxies.  In turn, this appears coherent with our image of rapid star formation in intermediate redshift galaxies.

\begin{acknowledgements}
NJRF acknowledges useful discussions with N. Abel on the CLOUDY capabilities to model molecular clouds. We thank the IRAM staff in Granada for their help with the observations. 
We thank John Cannon and  the SINGS team for the \emph{Spitzer} images. We also thank Fabian Walter for the use of the de Blok \& Walter \ion{H}{i} data.
\end{acknowledgements}

\bibliographystyle{aa}
\bibliography{11722.bib}

\end{document}